\documentclass[10pt]{article}
\usepackage[margin=1.0in]{geometry}

\usepackage{times}
\usepackage{latexsym}
\usepackage[T1]{fontenc}
\usepackage[utf8]{inputenc}
\usepackage{microtype}
\usepackage{inconsolata}
\usepackage{xcolor}
\usepackage[most]{tcolorbox}
\usepackage{pgfplots}
\usepackage{float}
\usepackage{tikz}
\usepackage{xparse}
\usepackage{xstring}
\pgfplotsset{compat=1.18}

\usepackage{amsmath}
\usepackage{pgf}
\usepackage{lmodern}
\usepackage{wasysym}
\usepackage[round]{natbib}

\usepackage{pdfpages}

\usepackage[utf8]{inputenc}
\usepackage[T1]{fontenc}
\usepackage{hyperref}
\usepackage{booktabs}
\usepackage{url}
\usepackage{amsfonts}
\usepackage{nicefrac}
\usepackage{microtype}
\usepackage{xcolor}

\usepackage{listings}
\usepackage{cleveref}

\usepackage{subcaption}
\usepackage{graphicx}
\usepackage{accents}
\usepackage{amsthm}
\usepackage{amssymb}
\usepackage{relsize}
\usepackage{tabularx}
\usepackage{thmtools, thm-restate}
\usepackage{xspace}
\usepackage{makecell}
\usepackage{xcolor, booktabs, colortbl}

\usepackage[dvipsnames]{xcolor}
\usepackage{soul}

\definecolor{clsA}{RGB}{230,159,0}
\definecolor{clsB}{RGB}{86,180,233}
\definecolor{clsC}{RGB}{0,158,115}
\definecolor{clsD}{RGB}{240,228,66}

\newcommand{\classA}[1]{
  \begingroup\sethlcolor{clsA!25}\hl{#1}\endgroup}
\newcommand{\classB}[1]{
  \begingroup\sethlcolor{clsB!25}\hl{#1}\endgroup}
\newcommand{\classC}[1]{
  \begingroup\sethlcolor{clsC!25}\hl{#1}\endgroup}
\newcommand{\classD}[1]{
  \begingroup\sethlcolor{clsD!25}\hl{#1}\endgroup}

\usepackage{eqparbox}

\input{survey_plot_macros}

\title{ML Researchers Support Openness in Peer Review\\But Are Concerned About Resubmission Bias}

\author{Vishisht Rao, Justin Payan, Andrew McCallum, and Nihar B. Shah \\{\normalsize \{vsrao, jpayan, nihars\}@cs.cmu.edu, mccallum@cs.umass.edu}\\{\normalsize Carnegie Mellon University, University of Massachusetts Amherst}}
\date{}

\begin{document}

\maketitle

\begin{abstract}
In recent years, peer-review venues have increasingly adopted \emph{open reviewing policies} that publicly release anonymized reviews and permit public commenting during or after the review process. Venues have adopted a wide variety of policies, and there is still ongoing debate about the benefits and drawbacks of each decision. 
To address this debate, we conduct a survey of $2{,}385$ reviewers, authors, and other peer-review participants to understand their experiences with and opinions of open reviewing policies. Our key findings are:

\begin{itemize}
    \item \textbf{Overall preferences:} Most respondents prefer at least some openness in the review process, with over $80\%$ of respondents supporting release of reviews for accepted papers and allowing public comments after acceptance. However, only $27.1\%$ of respondents support release of rejected manuscripts. 
    \item \textbf{Key perceived benefits of open reviewing:}
    \begin{itemize}
        \item Improvements in public understanding by providing more context about papers ($75.3\%$ of respondents) and educating novice reviewers ($57.8\%$ of respondents).
        \item The decision process is more fair ($56.6\%$ of respondents).
        \item Stronger incentives to write a high quality review ($48.0\%$ of respondents).
    \end{itemize}
    \item \textbf{Key perceived challenges of open reviewing:} 
    \begin{itemize}
        \item Potential for resubmission bias~\citep{stelmakh2021prior}, in which future reviewers are more likely to reject resubmissions when they can see a prior venue has rejected them (ranked top impact of open reviewing by $41\%$ of respondents, and mentioned in over $50\%$ of free responses).
        \item Fear of de-anonymization among reviewers ($33.2\%$ of respondents).
        \item Potential for noise and abuse in commenting (mentioned in $30$ free responses).
    \end{itemize}
    \item \textbf{AI and open peer review:} 
    \begin{itemize}
        \item Participants believe releasing rejected manuscripts can deter AI slop manuscript submissions ($71.9\%$ of respondents) and AI-generated reviews ($38.9\%$ of respondents).
        \item Respondents are split regarding peer-review venues generating official AI reviews, with $56.0\%$ opposed and $44.0\%$ supportive.
    \end{itemize}
\end{itemize} 
We also use AI to annotate $4{,}244$ open reviews from two top peer-review venues in computer science. We find that the fully open venue (ICLR) has higher levels of correctness and completeness than the partially open venue (NeurIPS). The effect size is small for correctness and very small for completeness, and both are statistically significant. We also find that there is no statistically significant difference in the level of substantiation.

We release the full set of anonymous survey responses along with the analysis code at \url{https://github.com/justinpayan/OpenReviewAnalysis}. 
\end{abstract}

\section{Introduction}

In recent years, the peer-review system has come under increased strain, leading  to numerous ideas to improve the process. As part of this effort, many venues have adopted \emph{open reviewing practices}, which open some parts of the peer-review process to the public, both during and after review. 
 
Proponents of these practices tout their benefits (e.g., increased transparency and community engagement), but open reviewing can also have negative impacts (e.g., resubmission bias and reduced willingness to participate as authors or reviewers). 

Open reviewing practices are increasingly widespread in prestigious publication venues, making it vital to understand the implications of these policies. The journal \emph{Nature} announced in June $2025$ that all of its reviews will be published anonymously along with published manuscripts \citep{nature2025transparent}. In computer science and machine learning, many peer-review venues host open processes on OpenReview.net, which was created in 2013 with the goal of enabling experimentation in peer review, by offering flexible configuration across a wide variety of reviewing workflows, including both traditional closed reviewing and open reviewing. Some of the largest and most prestigious machine learning venues have moved to OpenReview.net, including the Annual Conference on Neural Information Processing Systems (NeurIPS), the International Conference on Learning Representations (ICLR), and the International Conference on Machine Learning (ICML).\footnote{In computer science and machine learning, conferences review full-length papers, are typically ranked at par or higher than journals, and serve as a terminal publication venue.} \citet{soergel2013open} surveyed participants of ICLR after its first year on OpenReview ($2013$), finding broad support for many of the proposed benefits of open reviewing. It is not clear how the research community experiences these practices $12$ years later.

To fill this gap, we conduct an anonymous survey of researchers who have recently participated in prestigious, peer-reviewed machine learning conferences like NeurIPS, ICLR, and ICML. We ask survey respondents about their experiences as both authors and reviewers at these conferences, as well as their overall opinions about open policies. 
Our survey population is broadly representative of the machine learning community. We received $2385$ completed survey responses. The top demographic groups are PhD students (approximately $40\%$ of respondents), professors (approximately $25\%$ of respondents), and industry practitioners (approximately $20\%$ of respondents). We outline the key results of the survey in \Cref{subsec:keyobs}, and we provide a detailed analysis in \Cref{sec:survey_analysis}. Throughout the analysis, we discuss comparisons of our results with findings from closely-related surveys, paying special attention to that of \citet{soergel2013open}.

To supplement the survey results, we also use AI to compare publicly-available review text from ICLR $2023$ and NeurIPS $2022$. 
ICLR releases all of its submitted manuscripts and reviews to the public, including rejected manuscripts and their reviews. NeurIPS releases reviews for accepted papers, and allows authors to opt-in to release of rejected manuscripts and reviews. Since we cannot access closed reviews (which are not available to us for research \emph{by definition}),  comparison between two open venues with differing degrees of openness is the best available method to understand the impacts of open policies. We compare multiple metrics of review quality across the two conferences, including the fraction of incorrect statements, the coverage of the paper, and the fraction of negative judgments which are unsubstantiated. These results are presented in \Cref{sec:llm_evaluation}. We find a statistically significant effect showing that ICLR $2023$ reviews are approximately $6.5\%$ more correct and $2.2\%$ more complete on average than NeurIPS $2022$ reviews, with no significant differences in substantiation. 

We close with a discussion of related work in \Cref{sec:related_work}, and limitations and conclusions in \Cref{sec:limitations_and_conc}.

\subsection{Open reviewing policies under consideration}
\label{subsec:openreviewingpolicies}

We consider three commonly-used sets of reviewing policies, which we call \textbf{closed}, \textbf{partially open}, and \textbf{fully open} (nomenclature due to \citealp{yang2025paper}). An important point to note is that in all three systems, the identities of the reviewers are \textit{anonymized} to the public and to the authors.
Our survey asks about authors' and reviewers' opinions and experiences with these three systems of reviewing. 
The systems are:

\begin{itemize}
\item \textbf{Closed:} All reviews for a particular paper are visible only to the author, the reviewers assigned to that paper, and the meta-reviewer(s)\footnote{Meta-reviewers are members of the reviewing committee who compile the reports from reviewers and recommend final decisions. Two common levels of meta-reviewer are \emph{area chair} and \emph{senior area chair}.} assigned to that paper. Manuscripts are not released to the public, other than the camera-ready copies of accepted papers. Public commenting is never allowed. ICML used this model until $2025$.
\item \textbf{Partially open:} Reviews for accepted manuscripts are released to the public after the conference notification deadline, along with the accepted manuscripts themselves. Rejected manuscripts and their reviews are not released by default, though the authors of rejected manuscripts may opt in to releasing the manuscripts and reviews. Finally, the public may comment non-anonymously on public manuscripts and reviews after the notification deadline. NeurIPS uses this model.
\item \textbf{Fully open:} Reviews for all manuscripts are released to the public after the review submission deadline. All manuscripts are released upon submission, and rejected manuscripts remain visible. The public may comment anonymously or non-anonymously (they decide) on public manuscripts and reviews after their release. ICLR uses this model.
\end{itemize}

\subsection{Key observations}
\label{subsec:keyobs}
We now outline our main findings. Overall, we find \textbf{strong support for open reviewing policies}. Over 90\% of respondents prefer at least some openness in the review process. We asked respondents to select between closed, partially open, and fully open review release policies. We find that 33\% support releasing reviews of all submitted papers, and 56\% support releasing reviews for accepted papers with rejected papers' authors having the option to release their reviews (a total of 89\% supporting some openness). We also asked respondents to select between closed, partially open, and fully open commenting policies. We find that 38\% support public commenting anytime after the submission deadline, and 45\% support public commenting only after the paper acceptance decisions are released (a total of 84\% supporting some openness). When respondents select between closed, partially open, and fully open manuscript release policies, we find less support, approximately $27\%$, for releasing all rejected manuscripts. The most-mentioned topic in the free-response question at the end of the survey is explicit and enthusiastic support for open reviewing policies.

We list the main perceived benefits and challenges in open reviewing, along with an overview of how recent AI advancements interact with open reviewing practices:
\begin{itemize}
    \item \textbf{Perceived benefits of open reviewing policies}:
    \begin{itemize}
        \item \textbf{Improving public understanding}. Approximately 75\% of participants think public reviews and comments have been useful in understanding papers they did not write or review, and approximately 60\% say public reviews have helped them learn to write reviews. These effects are generally stronger for PhD students.
        \item \textbf{Incentive to produce higher quality reviews.} Approximately 50\% of respondents agree with the statement ``\emph{When a conference might make my review public, I feel more incentivized to write a high quality review}.'' In contrast, 15\% of respondents disagree and 35\% are neutral. 
        \item \textbf{More fair decision process.} Approximately 55\% of respondents think that open reviewing policies ``\emph{result in a decision process that is more fair},'' while 7\% say that the process is less fair, and the remaining 38\% are neutral.
    \end{itemize}
    \item \textbf{Perceived challenges for open reviewing policies}:
    \begin{itemize}
       \item \textbf{Resubmission bias}~\citep{stelmakh2021prior}. We asked respondents to rank impacts -- positive or negative -- of fully open reviewing policies that release rejected submissions and reviews, considering both how likely each impact is and how consequential it would be. The potential for resubmission bias (in which future reviewers are negatively biased against submissions rejected by a prior venue) was considered the most important overall impact, being ranked highest among five options by about $40\%$ of respondents. About $30\%$ of respondents say they sometimes refrain from submitting to fully open conferences. When asked why, about 50\% of those who replied ($235$ of $411$ respondents) say they refrain because of potential resubmission bias.
       \item \textbf{Fear of de-anonymization.}  Approximately one-third of reviewers are afraid they could be de-anonymized when conferences release reviews.
       \item \textbf{Noisy public comments.} A total of $30$ free text responses discuss aggressive/noisy public comments, especially during the decision process. People talk about modifying their reviews to avoid public fights, and they are concerned about targeted public campaigns for/against papers.
    \end{itemize}
    \item \textbf{AI and open peer review}:
    \begin{itemize}
        \item \textbf{Deterrence of AI reviews.} Approximately $40\%$ of respondents believe that open reviewing policies deter reviewers from submitting reviews wholly generated with AI, compared to $10\%$ who believe open reviewing policies encourage this behavior. 
        \item \textbf{Deterrence of AI submissions.} Approximately $70\%$ of respondents believe that releasing rejected manuscripts along with author names can deter AI slop submissions, and $45\%$ of respondents say they would prioritize deterrence of extremely poor-quality submissions over retention of more high-quality submissions. Approximately $40\%$ of respondents explicitly prefer retention of high-quality submissions even at the expense of admitting more low-quality submissions. 
        \item \textbf{Support for official AI reviews.} Slightly over half ($56\%$) of respondents do not support use of official AI-generated reviews for submissions, compared to $44\%$ who do support its use. When the latter group was asked who should be able to view these AI-generated reviews, the most popular response was ``\emph{General public, Authors, Meta-reviewers, Reviewers}'' (approximately $11\%$ of all respondents) and ``\emph{Authors, Meta-reviewers, Reviewers}'' (approximately $7\%$ of all respondents).
    \end{itemize}
\end{itemize}

We also compare our results with the outcomes of the survey conducted in 2013 by \citet{soergel2013open}. Overall, our respondents are still supportive of open reviewing policies, but with lower percentages than those reported by \citet{soergel2013open}.
\begin{itemize}
    \item \citet{soergel2013open} asked participants to compare fully open and closed review policies. They found that $80\%$ of respondents would recommend a fully open reviewing policy over a closed policy. It is unclear if their respondents would have preferred partially open policies if given the option. Our respondents support fully open policies at between $25$-$40\%$, depending on the policy. Support for partially or fully open policies is between $85$-$90\%$.
    \item Approximately half of respondents agree that open reviewing can incentivize higher quality reviewing, about the same as reported by \citet{soergel2013open} for a similar question.
    \item Reported experience of public comments helping to improve papers has decreased from approximately $60\%$ to approximately $30\%$.
    \item Belief that open reviewing makes the decision process more fair decreased from $75\%$ to approximately $55\%$.
\end{itemize}  

We release the survey and all de-identified data, along with all code used for analysis of the survey data and public reviews at \url{https://github.com/justinpayan/OpenReviewAnalysis}.\footnote{The survey was anonymous, and participants were asked to not include any confidential information in their responses. However, some demographic information (recent roles served in conferences,
primary affiliation type, and academic career level) could lead to de-anonymization. We de-identify this demographic data before release. Details of this process are explained in the repository.}

\section{Survey analysis}
\label{sec:survey_analysis}

We designed and deployed the survey using  the Qualtrics platform. 
Before deployment, we piloted the survey with a mix of graduate students and professors (details of the pilot are included in \Cref{appx:survey_pilot}). 
Although the survey is available in full online, all survey questions are repeated in \Cref{sec:survey_analysis}. The survey has $8$ sections in addition to the informed consent question: querying the roles served in recent top ML conferences, asking all participants for their experiences and opinions of open reviewing practices, $6$ author-only questions, $7$ reviewer-only questions, a section in which respondents rank the most important impacts of open reviewing practices, a section in which respondents express their opinions about what open reviewing practices conferences should implement, a free-response section, and a  demographic information section. It consists of $19$ multiple choice questions, $1$ question asking participants to rank $5$ options, and $3$ free text response questions. 

The ICLR $2025$ Program Chairs sent out our survey using an anonymous link through the OpenReview interface, to the list of email addresses for authors, reviewers, meta-reviewers, and organizers involved in ICLR $2025$. Participants were told the survey would close on September $5$, $2025$. We released the survey on Aug. $26$, $2025$ and closed it on Sept. $7$, $2025$. The survey was anonymous, and participants were asked to not include any identifiable information in their free-text responses.
The survey was approved by the CMU Institutional Review Board, study number $2024\_00000502$.

The remainder of this section is organized by the questions we answer throughout the survey:

\begin{itemize}
    \item \textbf{Who are the respondents?} (\Cref{subsec:whoarerespondents}) 
    \item \textbf{Do respondents find public reviews and comments useful?} (\Cref{subsec:dorespondentsfindpublic})
    \item \textbf{How do respondents view public commenting?} (\Cref{subsec:howdidrespondentsview})
    \item \textbf{Are respondents afraid of de-anonymization when reviewing?} (\Cref{subsec:wererespondentsafraid})
    \item \textbf{How does open reviewing impact the quality of reviews?} (\Cref{subsec:howdoesopenreviewingimpact})
    \item \textbf{Does open reviewing result in a more fair decision process?} (\Cref{subsec:doesopenreviewingresultin})
    \item \textbf{Does open reviewing compel extra experiments?} (\Cref{subsec:doesopenreviewingcompel})
    \item \textbf{Do participants refrain from submitting to fully open review, and why?} (\Cref{subsec:doparticipantsrefrain})
    \item \textbf{How does AI interact with open reviewing?} (\Cref{subsec:howdollms})
    \item \textbf{How do respondents rank potential impacts of open peer review?} (\Cref{subsec:howdorespondentsrank})
    \item \textbf{What else do respondents say in the free-response section?} (\Cref{subsec:whatelse})
    \item \textbf{How would participants design a future conference?} (\Cref{subsec:howwouldparticipantsdesign})
\end{itemize}

The final free-response question ``\emph{Do you have any additional thoughts you would like to express about open or closed reviewing practices?}'' is addressed explicitly in \Cref{subsec:whatelse}. However, we also reference free responses throughout our analysis, when relevant. We manually group these responses into common themes, and discard all responses which indicate only the ``None'' theme (e.g., responses like ``\emph{N/A}'' or ``\emph{I have nothing else to say}''). We refer to these themes when relevant.

\subsection{Who are the respondents?}
\label{subsec:whoarerespondents}

% Affiliation distribution (Multi-select: totals can exceed 100%)
% Options: Other, Government, Industry, Academia
% Total respondents: 2361

% We have three figures that we want to be subfigures of a single figure, stacked vertically.
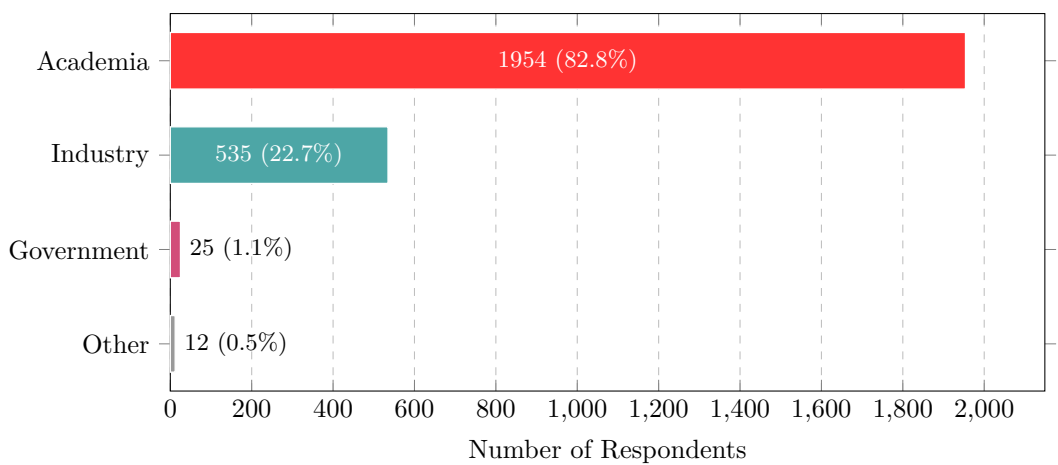
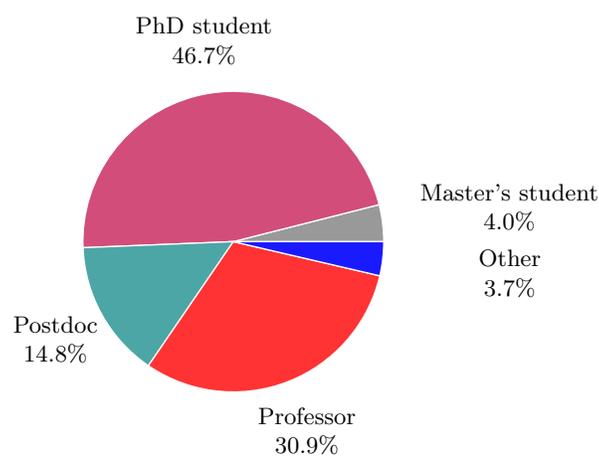
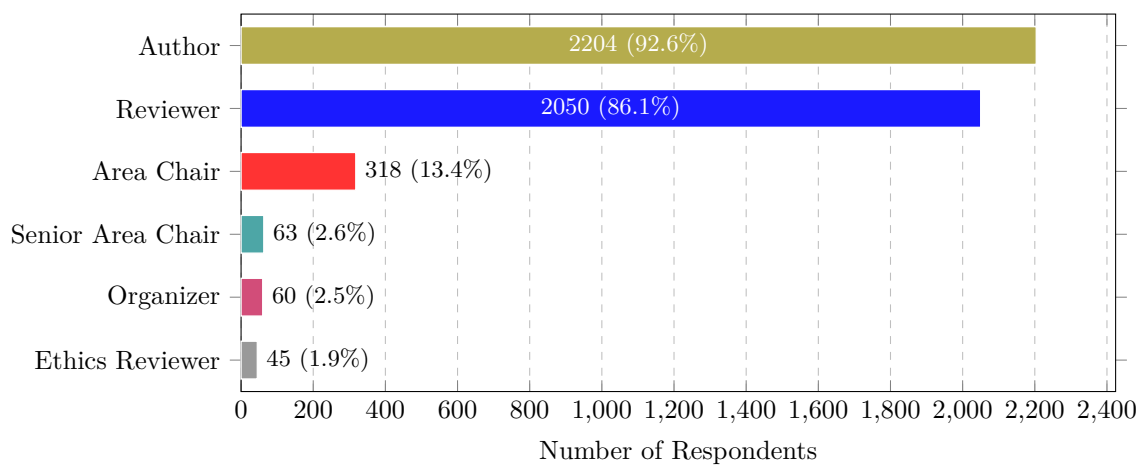
\begin{figure}
    \begin{subfigure}[t]{\textwidth}
    \centering
    \begin{tikzpicture}
    \begin{axis}[
        xbar,
        xmin=0,
        xmax=2149,
        width=0.8\textwidth,
        height=0.4\textwidth,
        xlabel={Number of Respondents},
        xlabel style={font=\normalsize},
        yticklabels={Other,Government,Industry,Academia},
        ytick={0,1,2,3},
        ymin=-0.5,
        ymax=3.5,
        bar width=8pt,
        bar shift=0pt,
        xmajorgrids=true,
        xminorgrids=true,
        grid style={dashed},
        clip=false
    ]
    
        \draw[fill=gray!80, draw=white, line width=0.5pt] (axis cs:0,-0.3) rectangle (axis cs:12,0.3);
        \node[font=\small, text=black, anchor=west] at (axis cs:12, 0) {12 (0.5\%)};
        \draw[fill=purple!70, draw=white, line width=0.5pt] (axis cs:0,0.7) rectangle (axis cs:25,1.3);
        \node[font=\small, text=black, anchor=west] at (axis cs:25, 1) {25 (1.1\%)};
        \draw[fill=teal!70, draw=white, line width=0.5pt] (axis cs:0,1.7) rectangle (axis cs:535,2.3);
        \node[font=\small, text=white] at (axis cs:267.5, 2) {535 (22.7\%)};
        \draw[fill=red!80, draw=white, line width=0.5pt] (axis cs:0,2.7) rectangle (axis cs:1954,3.3);
        \node[font=\small, text=white] at (axis cs:977.0, 3) {1954 (82.8\%)};
    
    \end{axis}
    \end{tikzpicture}
    \caption{Respondent affiliations, multiple selections allowed. ($n=2361$)}
    \label{fig:affiliations}
    \end{subfigure}
    \vspace{0.2cm}

    % Academic Role Pie Chart
    \begin{subfigure}[t]{\textwidth}
    \centering
    \begin{tikzpicture}[scale=2]
        \filldraw[fill=gray!80, draw=white, line width=0.5pt] (0,0) -- (0.00:1) arc (0.00:14.25:1) -- cycle;
        \filldraw[fill=purple!70, draw=white, line width=0.5pt] (0,0) -- (14.25:1) arc (14.25:182.31:1) -- cycle;
        \filldraw[fill=teal!70, draw=white, line width=0.5pt] (0,0) -- (182.31:1) arc (182.31:235.62:1) -- cycle;
        \filldraw[fill=red!80, draw=white, line width=0.5pt] (0,0) -- (235.62:1) arc (235.62:346.86:1) -- cycle;
        \filldraw[fill=blue!90, draw=white, line width=0.5pt] (0,0) -- (346.86:1) arc (346.86:360.00:1) -- cycle;
        \node[font=\small, align=center] at (7.13:1.85) {Master's student\\ 4.0\%};
        \node[font=\small, align=center] at (98.28:1.35) {PhD student\\ 46.7\%};
        \node[font=\small, align=center] at (208.97:1.35) {Postdoc\\ 14.8\%};
        \node[font=\small, align=center] at (291.24:1.35) {Professor\\ 30.9\%};
        \node[font=\small, align=center] at (353.43:1.85) {Other\\ 3.7\%};
    \end{tikzpicture}
     \caption{Academic roles of respondents who selected academia affiliation. ($n=1945$)}
    \label{fig:academic_roles}
    \end{subfigure}

    % Question 1: Roles Served at Conferences (Multi-select: totals can exceed 100%)
    % Total respondents: 2381
    \vspace{0.2cm}

    \begin{subfigure}[t]{\textwidth}
    \centering
    \begin{tikzpicture}
    \begin{axis}[
        xbar,
        xmin=0,
        xmax=2424,
        width=0.8\textwidth,
        height=0.4\textwidth,
        xlabel={Number of Respondents},
        xlabel style={font=\normalsize},
        yticklabels={Ethics Reviewer,Organizer,Senior Area Chair,Area Chair,Reviewer,Author},
        ytick={0,1,2,3,4,5},
        ymin=-0.5,
        ymax=5.5,
        bar width=8pt,
        bar shift=0pt,
        xmajorgrids=true,
        xminorgrids=true,
        grid style={dashed},
        clip=false
    ]
    
        \draw[fill=gray!80, draw=white, line width=0.5pt] (axis cs:0,-0.3) rectangle (axis cs:45,0.3);
        \node[font=\small, text=black, anchor=west] at (axis cs:45, 0) {45 (1.9\%)};
        \draw[fill=purple!70, draw=white, line width=0.5pt] (axis cs:0,0.7) rectangle (axis cs:60,1.3);
        \node[font=\small, text=black, anchor=west] at (axis cs:60, 1) {60 (2.5\%)};
        \draw[fill=teal!70, draw=white, line width=0.5pt] (axis cs:0,1.7) rectangle (axis cs:63,2.3);
        \node[font=\small, text=black, anchor=west] at (axis cs:63, 2) {63 (2.6\%)};
        \draw[fill=red!80, draw=white, line width=0.5pt] (axis cs:0,2.7) rectangle (axis cs:318,3.3);
        \node[font=\small, text=black, anchor=west] at (axis cs:318, 3) {318 (13.4\%)};
        \draw[fill=blue!90, draw=white, line width=0.5pt] (axis cs:0,3.7) rectangle (axis cs:2050,4.3);
        \node[font=\small, text=white] at (axis cs:1025.0, 4) {2050 (86.1\%)};
        \draw[fill=olive!70, draw=white, line width=0.5pt] (axis cs:0,4.7) rectangle (axis cs:2204,5.3);
        \node[font=\small, text=white] at (axis cs:1102.0, 5) {2204 (92.6\%)};
    
    \end{axis}
    \end{tikzpicture}
    \caption{Roles served at recent top machine learning conferences, multiple selections allowed. ($n=2381$)}
    \label{fig:question1_roles}
    \end{subfigure}
    \caption{Demographic information about survey respondents.}
    \label{fig:demo_info}
 \end{figure}

We received a total of $4102$ survey responses, of which $2385$ were marked as ``Finished'' by Qualtrics. Among the $1717$ respondents who did not complete the survey, only $153$ completed more than the first five questions. We also observe that inclusion of those who did not finish the survey does not impact any of the response distributions by more than $3\%$. We therefore only consider responses which are marked as ``Finished'' by Qualtrics for our analysis.

Demographic information is shown in \Cref{fig:demo_info}.  \Cref{fig:affiliations} shows that our respondents are $82.8\%$ academics, and $22.7\%$ from industry. Approximately half of academic respondents are PhD students, one-third are professors, and one-seventh are postdocs (\Cref{fig:academic_roles}). Only $1.0\%$ of respondents are undergraduate students. We group undergraduates with the respondents who select ``\emph{Other}'' in \Cref{fig:academic_roles}. We asked ``\emph{Which roles have you served in recent top machine learning conferences (including, but not limited to: ICLR, NeurIPS, and ICML)? Check all that apply.}'' Respondents selected from ``\emph{Author},'' ``\emph{Reviewer},'' ``\emph{Area Chair},'' ``\emph{Senior Area Chair},'' ``\emph{Organizer},'' and ``\emph{Ethics Reviewer}.'' The results are shown in \Cref{fig:question1_roles}.  A large majority of respondents have been both author and reviewer: $320$ respondents ($13.4\%$) select author but not reviewer, $168$ ($7.1\%$) select reviewer but not author, and $1882$ ($79.0\%$) have been both author and reviewer. Approximately $15\%$ of respondents have served as area chairs recently. We also received over $100$ responses from more senior respondents (senior area chairs and organizers).

\begin{figure}[H]
\centering
\hspace*{-2cm}
\begin{tikzpicture}
\begin{axis}[
    xbar,
    xmin=-90,
    xmax=90,
    width=0.7\textwidth,
    height=0.6\textwidth,
    xlabel={Percentage of Respondents},
    xlabel style={font=\normalsize},
    ytick=\empty,
    ymin=-0.5,
    ymax=5.5,
    bar width=8pt,
    bar shift=0pt,
    xmajorgrids=true,
    xminorgrids=true,
    ymajorgrids=true,
    grid style={dashed},
    grid style={dashed,
            dash pattern=on 4pt off 4pt,
            dash phase=2pt,
            color=gray!60},
    extra x ticks={0},
    extra x tick style={grid=major, major grid style={line width=1pt, draw=black}},
    x tick label style={/pgf/number format/assume math mode=true},
    xtick={-80,-60,-40,-20,0,20,40,60,80},
    xticklabels={-80,-60,-40,-20,0,20,40,60,80},
    clip=false,
    legend style={at={(0.5,1.02)}, anchor=south, legend columns=5, font=\small}
]

    \draw[color=gray!60, line width=0.4pt] (axis cs:-90,0) -- (axis cs:90,0);
    \draw[fill=blue!90, draw=white, line width=0.5pt] (axis cs:-48.17,-0.3) rectangle (axis cs:-42.74,0.3);
    \node[font=\tiny, text=white, anchor=center] at (axis cs:-45.46, 0) {5\%};
    \draw[fill=cyan!70, draw=white, line width=0.5pt] (axis cs:-42.74,-0.3) rectangle (axis cs:-19.00,0.3);
    \node[font=\small, text=white, anchor=center] at (axis cs:-30.87, 0) {23.7\%};
    \draw[fill=gray!80, draw=white, line width=0.5pt] (axis cs:-19.00,-0.3) rectangle (axis cs:19.00,0.3);
    \node[font=\small, text=white, anchor=center] at (axis cs:0.00, 0) {38.0\%};
    \draw[fill=orange!70, draw=white, line width=0.5pt] (axis cs:19.00,-0.3) rectangle (axis cs:44.20,0.3);
    \node[font=\small, text=white, anchor=center] at (axis cs:31.60, 0) {25.2\%};
    \draw[fill=red!90, draw=white, line width=0.5pt] (axis cs:44.20,-0.3) rectangle (axis cs:51.83,0.3);
    \node[font=\tiny, text=white, anchor=center] at (axis cs:48.01, 0) {8\%};
    \draw[color=gray!60, line width=0.4pt] (axis cs:-90,1) -- (axis cs:90,1);
    \draw[fill=blue!90, draw=white, line width=0.5pt] (axis cs:-34.43,0.7) rectangle (axis cs:-29.88,1.3);
    \node[font=\tiny, text=white, anchor=center] at (axis cs:-32.15, 1) {5\%};
    \draw[fill=cyan!70, draw=white, line width=0.5pt] (axis cs:-29.88,0.7) rectangle (axis cs:-17.60,1.3);
    \node[font=\tiny, text=white, anchor=center] at (axis cs:-23.74, 1) {12.3\%};
    \draw[fill=gray!80, draw=white, line width=0.5pt] (axis cs:-17.60,0.7) rectangle (axis cs:17.60,1.3);
    \node[font=\small, text=white, anchor=center] at (axis cs:0.00, 1) {35.2\%};
    \draw[fill=orange!70, draw=white, line width=0.5pt] (axis cs:17.60,0.7) rectangle (axis cs:55.99,1.3);
    \node[font=\small, text=white, anchor=center] at (axis cs:36.80, 1) {38.4\%};
    \draw[fill=red!90, draw=white, line width=0.5pt] (axis cs:55.99,0.7) rectangle (axis cs:65.57,1.3);
    \node[font=\tiny, text=white, anchor=center] at (axis cs:60.78, 1) {9.6\%};
    \draw[color=gray!60, line width=0.4pt] (axis cs:-90,2) -- (axis cs:90,2);
    \draw[fill=blue!90, draw=white, line width=0.5pt] (axis cs:-55.94,1.7) rectangle (axis cs:-43.67,2.3);
    \node[font=\tiny, text=white, anchor=center] at (axis cs:-49.80, 2) {12.3\%};
    \draw[fill=cyan!70, draw=white, line width=0.5pt] (axis cs:-43.67,1.7) rectangle (axis cs:-10.86,2.3);
    \node[font=\small, text=white, anchor=center] at (axis cs:-27.26, 2) {32.8\%};
    \draw[fill=gray!80, draw=white, line width=0.5pt] (axis cs:-10.86,1.7) rectangle (axis cs:10.86,2.3);
    \node[font=\small, text=white, anchor=center] at (axis cs:0.00, 2) {21.7\%};
    \draw[fill=orange!70, draw=white, line width=0.5pt] (axis cs:10.86,1.7) rectangle (axis cs:36.04,2.3);
    \node[font=\small, text=white, anchor=center] at (axis cs:23.45, 2) {25.2\%};
    \draw[fill=red!90, draw=white, line width=0.5pt] (axis cs:36.04,1.7) rectangle (axis cs:44.06,2.3);
    \node[font=\tiny, text=white, anchor=center] at (axis cs:40.05, 2) {8\%};
    \draw[color=gray!60, line width=0.4pt] (axis cs:-90,3) -- (axis cs:90,3);
    \draw[fill=blue!90, draw=white, line width=0.5pt] (axis cs:-52.06,2.7) rectangle (axis cs:-41.63,3.3);
    \node[font=\tiny, text=white, anchor=center] at (axis cs:-46.84, 3) {10.4\%};
    \draw[fill=cyan!70, draw=white, line width=0.5pt] (axis cs:-41.63,2.7) rectangle (axis cs:-21.01,3.3);
    \node[font=\small, text=white, anchor=center] at (axis cs:-31.32, 3) {20.6\%};
    \draw[fill=gray!80, draw=white, line width=0.5pt] (axis cs:-21.01,2.7) rectangle (axis cs:21.01,3.3);
    \node[font=\small, text=white, anchor=center] at (axis cs:0.00, 3) {42\%};
    \draw[fill=orange!70, draw=white, line width=0.5pt] (axis cs:21.01,2.7) rectangle (axis cs:42.36,3.3);
    \node[font=\small, text=white, anchor=center] at (axis cs:31.69, 3) {21.4\%};
    \draw[fill=red!90, draw=white, line width=0.5pt] (axis cs:42.36,2.7) rectangle (axis cs:47.94,3.3);
    \node[font=\tiny, text=white, anchor=center] at (axis cs:45.15, 3) {6\%};
    \draw[color=gray!60, line width=0.4pt] (axis cs:-90,4) -- (axis cs:90,4);
    \draw[fill=blue!90, draw=white, line width=0.5pt] (axis cs:-31.76,3.7) rectangle (axis cs:-25.16,4.3);
    \node[font=\tiny, text=white, anchor=center] at (axis cs:-28.46, 4) {7\%};
    \draw[fill=cyan!70, draw=white, line width=0.5pt] (axis cs:-25.16,3.7) rectangle (axis cs:-10.44,4.3);
    \node[font=\small, text=white, anchor=center] at (axis cs:-17.80, 4) {14.7\%};
    \draw[fill=gray!80, draw=white, line width=0.5pt] (axis cs:-10.44,3.7) rectangle (axis cs:10.44,4.3);
    \node[font=\small, text=white, anchor=center] at (axis cs:0.00, 4) {20.9\%};
    \draw[fill=orange!70, draw=white, line width=0.5pt] (axis cs:10.44,3.7) rectangle (axis cs:50.49,4.3);
    \node[font=\small, text=white, anchor=center] at (axis cs:30.46, 4) {40\%};
    \draw[fill=red!90, draw=white, line width=0.5pt] (axis cs:50.49,3.7) rectangle (axis cs:68.24,4.3);
    \node[font=\small, text=white, anchor=center] at (axis cs:59.36, 4) {17.8\%};
    \draw[color=gray!60, line width=0.4pt] (axis cs:-90,5) -- (axis cs:90,5);
    \draw[fill=blue!90, draw=white, line width=0.5pt] (axis cs:-17.69,4.7) rectangle (axis cs:-13.66,5.3);
    \node[font=\tiny, text=white, anchor=center] at (axis cs:-15.67, 5) {4\%};
    \draw[fill=cyan!70, draw=white, line width=0.5pt] (axis cs:-13.66,4.7) rectangle (axis cs:-7.06,5.3);
    \node[font=\tiny, text=white, anchor=center] at (axis cs:-10.36, 5) {7\%};
    \draw[fill=gray!80, draw=white, line width=0.5pt] (axis cs:-7.06,4.7) rectangle (axis cs:7.06,5.3);
    \node[font=\small, text=white, anchor=center] at (axis cs:0.00, 5) {14.1\%};
    \draw[fill=orange!70, draw=white, line width=0.5pt] (axis cs:7.06,4.7) rectangle (axis cs:54.16,5.3);
    \node[font=\small, text=white, anchor=center] at (axis cs:30.61, 5) {47.1\%};
    \draw[fill=red!90, draw=white, line width=0.5pt] (axis cs:54.16,4.7) rectangle (axis cs:82.31,5.3);
    \node[font=\small, text=white, anchor=center] at (axis cs:68.24, 5) {28.2\%};

    \addlegendimage{area legend, fill=blue!90, draw=white, line width=0.5pt}
    \addlegendentry{Strongly disagree}
    \addlegendimage{area legend, fill=cyan!70, draw=white, line width=0.5pt}
    \addlegendentry{Disagree}
    \addlegendimage{area legend, fill=gray!80, draw=white, line width=0.5pt}
    \addlegendentry{Neither agree nor disagree}
    \addlegendimage{area legend, fill=orange!70, draw=white, line width=0.5pt}
    \addlegendentry{Agree}
    \addlegendimage{area legend, fill=red!90, draw=white, line width=0.5pt}
    \addlegendentry{Strongly agree}

    \node[font=\normalsize, align=right, text width=6cm, anchor=east] at (axis cs:-90, 0) {vi. Compelled to complete\\ unnecessary analyses};
    \node[font=\normalsize, align=right, text width=6cm, anchor=east] at (axis cs:-90, 1) {v. More incentivized to\\ write quality reviews};
    \node[font=\normalsize, align=right, text width=6cm, anchor=east] at (axis cs:-90, 2) {iv. Afraid of de-anonymization\\ as reviewer};
    \node[font=\normalsize, align=right, text width=6cm, anchor=east] at (axis cs:-90, 3) {iii. Comments have\\ helped improve papers};
    \node[font=\normalsize, align=right, text width=6cm, anchor=east] at (axis cs:-90, 4) {ii. Learned to write reviews\\ from public reviews};
    \node[font=\normalsize, align=right, text width=6cm, anchor=east] at (axis cs:-90, 5) {i. Reviews or comments\\ useful when viewing papers};

    \node[font=\small, align=left, anchor=west] at (axis cs:92, 0) {(n=2190)};
    \node[font=\small, align=left, anchor=west] at (axis cs:92, 1) {(n=2045)};
    \node[font=\small, align=left, anchor=west] at (axis cs:92, 2) {(n=2045)};
    \node[font=\small, align=left, anchor=west] at (axis cs:92, 3) {(n=2187)};
    \node[font=\small, align=left, anchor=west] at (axis cs:92, 4) {(n=2045)};
    \node[font=\small, align=left, anchor=west] at (axis cs:92, 5) {(n=2380)};

\end{axis}
\end{tikzpicture}
\caption{Agreement with statements about open reviewing practices.}
\label{fig:combined_likert}
\end{figure}
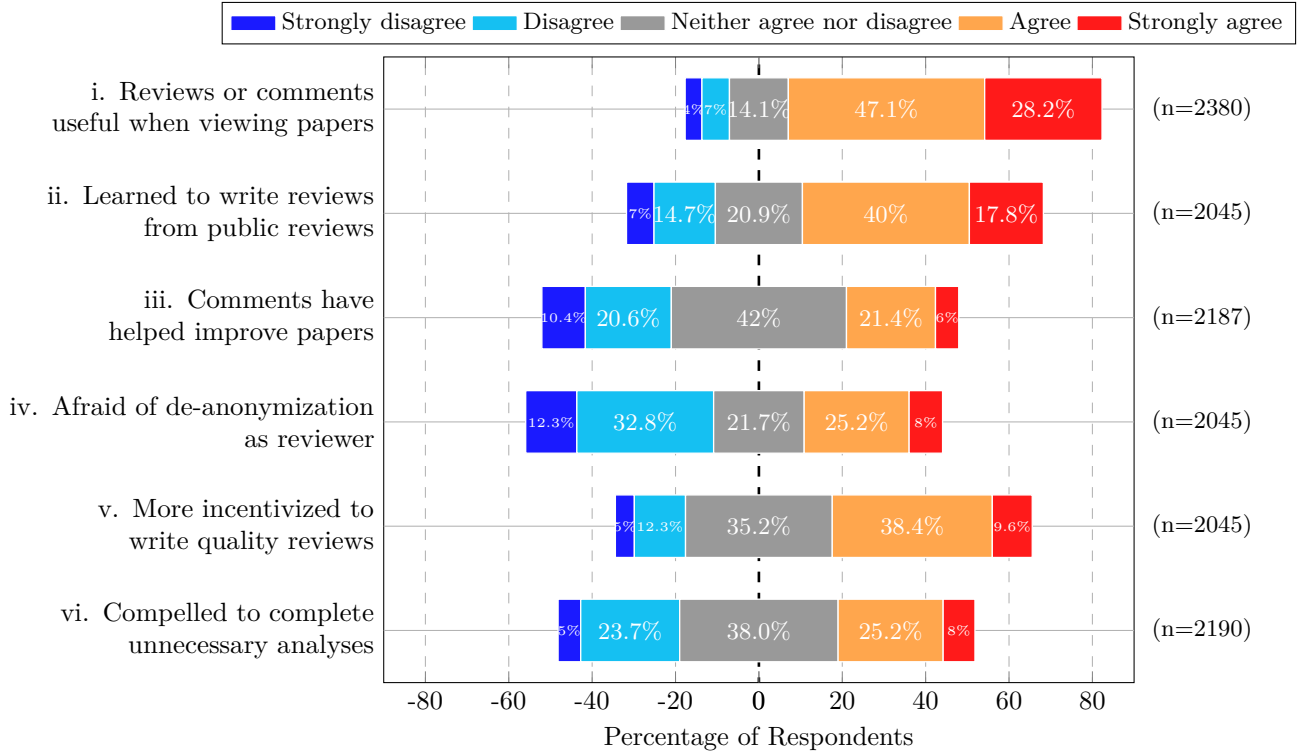

\subsection{Do respondents find public reviews and comments useful?}
\label{subsec:dorespondentsfindpublic}

We asked respondents to rate their agreement with the statement ``\emph{When viewing papers for which I am neither an author nor a reviewer, I sometimes find publicly-available, official reviews or comments useful}.'' Respondents indicated their agreement on a $5$-point Likert scale with options ``\emph{Strongly disagree},'' ``\emph{Disagree},'' ``\emph{Neither agree nor disagree},'' ``\emph{Agree},'' and ``\emph{Strongly agree}.'' 
Row (i) of \Cref{fig:combined_likert} shows that approximately $75\%$ of participants agree or strongly agree with this statement. We also break agreement down by career stage in \Cref{fig:career_stage_breakdown_2}. We see that agreement lies within $70$-$80\%$ for all groups, but that the level of agreement marginally decreases with seniority.

\careerStageBreakdown{Agreement with the statement ``\emph{When viewing papers for which I am neither an author nor a reviewer, I sometimes find publicly-available, official reviews or comments useful},'' broken down by career stage. ($n=2053$)}{fig:career_stage_breakdown_2}{830,272,565,386,4,4,13.6,45.9,32.9,2,7.7,11.4,49.3,29.4,4,9.4,15.4,48.1,22.7,5,7.8,15.0,47.9,24.4}{0.3}

We asked participants to rate their agreement with the statement ``\emph{Publicly-available, official reviews have helped me learn how to write reviews}.'' Row (ii) in \Cref{fig:combined_likert} shows approximately  $60\%$ agree or strongly agree, and approximately $20\%$ disagree or strongly disagree. We break agreement down by career stage in \Cref{fig:career_stage_breakdown_11}. Among professors, approximately $40\%$ of respondents agree with this statement, while approximately $70\%$ of PhD students agree.
Overall, these responses suggest that public reviews and comments are informative for the community, and this effect is strongest for more junior community members.

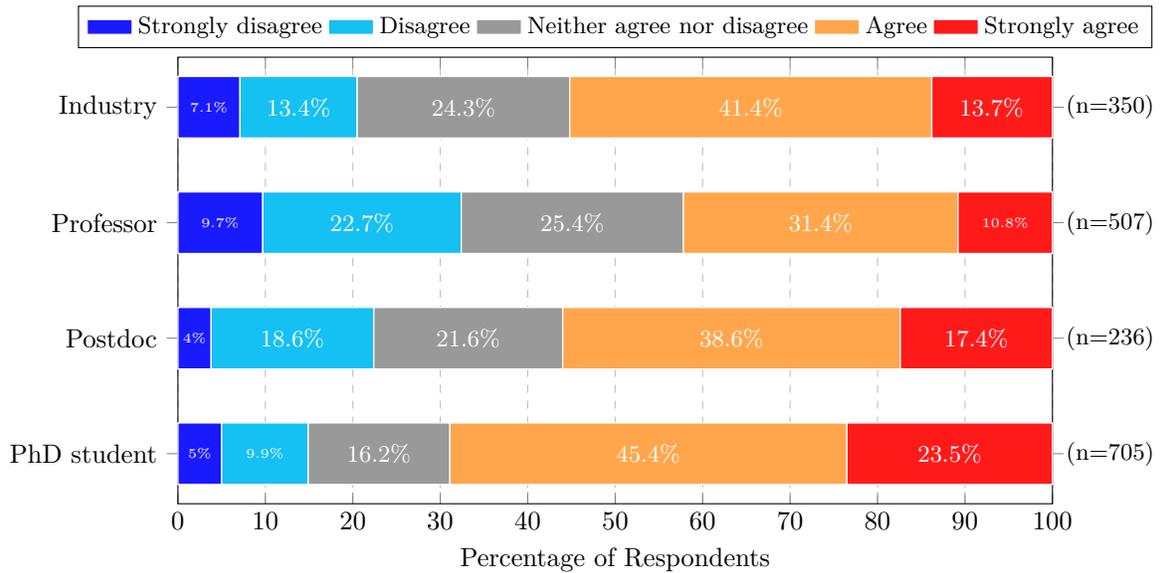
\careerStageBreakdown{Agreement with the statement ``\emph{Publicly-available, official reviews have helped me learn how to write reviews},'' broken down by career stage. ($n=1798$)}{fig:career_stage_breakdown_11}{705,236,507,350,5.0,9.9,16.2,45.4,23.5,3.8,18.6,21.6,38.6,17.4,9.7,22.7,25.4,31.4,10.8,7.1,13.4,24.3,41.4,13.7}{0.3}

\subsection{How do respondents view public commenting?}
\label{subsec:howdidrespondentsview}

We asked those who had been authors of papers in recent years to rate their agreement with the statement ``\emph{Insights from commenters other than anonymous reviewers have helped me to improve a paper I submitted}.'' The result is shown in row (iii) of \Cref{fig:combined_likert}. We see that approximately $30\%$ of respondents agree, and approximately $30\%$ disagree. In $2013$, approximately $60\%$ agreed and $10\%$ disagreed with a similar statement \citep{soergel2013open}. This suggests that insights from public comments are still useful (with about one-third of respondents agreeing), but much less overwhelmingly useful than they appeared in $2013$. 

\begin{figure}
\centering
\begin{subfigure}{0.9\textwidth}
\centering
\vspace{0.2cm}
    \begin{tikzpicture}
    \begin{axis}[
        xbar,
        xmin=0,
        xmax=100,
        width=0.75\textwidth,
        height=0.175\textwidth,
        xlabel={Percentage of Respondents},
        xlabel style={font=\normalsize},
        yticklabels={Comfort commenting},
        ytick={0},
        ymin=-0.25,
        ymax=0.25,
        bar width=6pt,
        bar shift=0pt,
        xmajorgrids=true,
        xminorgrids=true,
        grid style={dashed},
        clip=false,
        legend style={at={(0.5,1.15)}, anchor=south, legend columns=3, font=\small}
    ]
    
    % Define bar dimensions (barheight variable for easy adjustment)
    \def\barheight{0.2}
    \pgfmathsetmacro{\yBar}{0}
    \pgfmathsetmacro{\yBarBottom}{\yBar-\barheight}
    \pgfmathsetmacro{\yBarTop}{\yBar+\barheight}
    
    % Create stacked bar segments
    \draw[fill=gray!80, draw=white, line width=0.5pt] (axis cs:0, \yBarBottom) rectangle (axis cs:32.04047217537943, \yBarTop);
    \node[font=\small, text=white] at (axis cs:16.020236087689714, \yBar) {{32.0\%}};
    \draw[fill=purple!70, draw=white, line width=0.5pt] (axis cs:32.04047217537943, \yBarBottom) rectangle (axis cs:76.01180438448566, \yBarTop);
    \node[font=\small, text=white] at (axis cs:54.02613827993254, \yBar) {{44.0\%}};
    \draw[fill=teal!70, draw=white, line width=0.5pt] (axis cs:76.01180438448566, \yBarBottom) rectangle (axis cs:100.0, \yBarTop);
    \node[font=\small, text=white] at (axis cs:88.00590219224283, \yBar) {{24.0\%}};
    \addlegendimage{area legend, fill=gray!80, draw=white, line width=0.5pt}
    \addlegendentry{Not comfortable}
    \addlegendimage{area legend, fill=purple!70, draw=white, line width=0.5pt}
    \addlegendentry{Only anonymously}
    \addlegendimage{area legend, fill=teal!70, draw=white, line width=0.5pt}
    \addlegendentry{Comfortable non-anonymously}
    \end{axis}
    \end{tikzpicture}
\caption{Distribution of responses to ``\emph{In review settings where public commenting is allowed, how comfortable are you commenting on papers for which you are neither an author nor a reviewer?}'' (n=2372)}
\label{fig:question3_combined}
\vspace{0.3cm}
\end{subfigure}

\vspace{0.5cm}

\begin{subfigure}[t]{0.9\textwidth}
\centering
    \begin{tikzpicture}
    \begin{axis}[
        xbar,
        xmin=0,
        xmax=100,
        width=0.8\textwidth,
        height=0.35\textwidth,
        xlabel={Percentage of Respondents},
        xlabel style={font=\normalsize},
        yticklabels={Not Comfortable,Anonymous Only,Non-Anonymous},
        ytick={0.0,0.4,0.8},
        ymin=-0.175,
        ymax=0.9750000000000001,
        bar width=6pt,
        bar shift=0pt,
        xmajorgrids=true,
        xminorgrids=true,
        grid style={dashed},
        clip=false,
        legend style={at={(0.5,1.02)}, anchor=south, legend columns=5, font=\small}
    ]
    
    % Define barheight variable for easy adjustment (change bar_height in Python code)
    % Current value: 0.125 (bar height is half-height from center)
    
    % Define y-positions for each bar
    \pgfmathsetmacro{\yNotComfortable}{0.0}
    \pgfmathsetmacro{\yAnonymousOnly}{0.4}
    \pgfmathsetmacro{\yNonAnonymous}{0.8}
    
    % Not Comfortable bar
    \pgfmathsetmacro{\yNotComfortableBottom}{\yNotComfortable-0.125}
    \pgfmathsetmacro{\yNotComfortableTop}{\yNotComfortable+0.125}
    \draw[fill=blue!90, draw=white, line width=0.5pt] (axis cs:0, \yNotComfortableBottom) rectangle (axis cs:6.3, \yNotComfortableTop);
    \pgfmathparse{6.3>=13}
    \ifnum\pgfmathresult=1
        \node[font=\small, text=white] at (axis cs:{6.3/2}, \yNotComfortable) {\formatPctLabel{6.3}\%};
    \else
        \pgfmathparse{6.3>0}
        \ifnum\pgfmathresult=1 \node[font=\tiny, text=white] at (axis cs:{6.3/2}, \yNotComfortable) {\formatPctLabel{6.3}\%}; \fi
    \fi
    \draw[fill=cyan!70, draw=white, line width=0.5pt] (axis cs:6.3, \yNotComfortableBottom) rectangle (axis cs:15.899999999999999, \yNotComfortableTop);
    \pgfmathparse{9.6>=13}
    \ifnum\pgfmathresult=1
        \node[font=\small, text=white] at (axis cs:{6.3+9.6/2}, \yNotComfortable) {\formatPctLabel{9.6}\%};
    \else
        \pgfmathparse{9.6>0}
        \ifnum\pgfmathresult=1 \node[font=\tiny, text=white] at (axis cs:{6.3+9.6/2}, \yNotComfortable) {\formatPctLabel{9.6}\%}; \fi
    \fi
    \draw[fill=gray!80, draw=white, line width=0.5pt] (axis cs:15.899999999999999, \yNotComfortableBottom) rectangle (axis cs:31.799999999999997, \yNotComfortableTop);
    \pgfmathparse{15.9>=13}
    \ifnum\pgfmathresult=1
        \node[font=\small, text=white] at (axis cs:{15.899999999999999+15.9/2}, \yNotComfortable) {\formatPctLabel{15.9}\%};
    \else
        \pgfmathparse{15.9>0}
        \ifnum\pgfmathresult=1 \node[font=\tiny, text=white] at (axis cs:{15.899999999999999+15.9/2}, \yNotComfortable) {\formatPctLabel{15.9}\%}; \fi
    \fi
    \draw[fill=orange!70, draw=white, line width=0.5pt] (axis cs:31.799999999999997, \yNotComfortableBottom) rectangle (axis cs:78.19999999999999, \yNotComfortableTop);
    \pgfmathparse{46.4>=13}
    \ifnum\pgfmathresult=1
        \node[font=\small, text=white] at (axis cs:{31.799999999999997+46.4/2}, \yNotComfortable) {\formatPctLabel{46.4}\%};
    \else
        \pgfmathparse{46.4>0}
        \ifnum\pgfmathresult=1 \node[font=\tiny, text=white] at (axis cs:{31.799999999999997+46.4/2}, \yNotComfortable) {\formatPctLabel{46.4}\%}; \fi
    \fi
    \draw[fill=red!90, draw=white, line width=0.5pt] (axis cs:78.19999999999999, \yNotComfortableBottom) rectangle (axis cs:100, \yNotComfortableTop);
    \pgfmathparse{21.7>=13}
    \ifnum\pgfmathresult=1
        \node[font=\small, text=white] at (axis cs:{78.19999999999999+21.7/2}, \yNotComfortable) {\formatPctLabel{21.7}\%};
    \else
        \pgfmathparse{21.7>0}
        \ifnum\pgfmathresult=1 \node[font=\tiny, text=white] at (axis cs:{78.19999999999999+21.7/2}, \yNotComfortable) {\formatPctLabel{21.7}\%}; \fi
    \fi

    % Anonymous Only bar
    \pgfmathsetmacro{\yAnonymousOnlyBottom}{\yAnonymousOnly-0.125}
    \pgfmathsetmacro{\yAnonymousOnlyTop}{\yAnonymousOnly+0.125}
    \draw[fill=blue!90, draw=white, line width=0.5pt] (axis cs:0, \yAnonymousOnlyBottom) rectangle (axis cs:3.2, \yAnonymousOnlyTop);
    \node[font=\tiny, text=white] at (axis cs:1.6, \yAnonymousOnly) {{3\%}};
    \draw[fill=cyan!70, draw=white, line width=0.5pt] (axis cs:3.2, \yAnonymousOnlyBottom) rectangle (axis cs:8.8, \yAnonymousOnlyTop);
    \node[font=\tiny, text=white] at (axis cs:6.0, \yAnonymousOnly) {{5.6\%}};
    \draw[fill=gray!80, draw=white, line width=0.5pt] (axis cs:8.8, \yAnonymousOnlyBottom) rectangle (axis cs:23.5, \yAnonymousOnlyTop);
    \node[font=\small, text=white] at (axis cs:16.15, \yAnonymousOnly) {{14.7\%}};
    \draw[fill=orange!70, draw=white, line width=0.5pt] (axis cs:23.5, \yAnonymousOnlyBottom) rectangle (axis cs:72.4, \yAnonymousOnlyTop);
    \node[font=\small, text=white] at (axis cs:47.95, \yAnonymousOnly) {{48.9\%}};
    \draw[fill=red!90, draw=white, line width=0.5pt] (axis cs:72.4, \yAnonymousOnlyBottom) rectangle (axis cs:100.10000000000001, \yAnonymousOnlyTop);
    \node[font=\small, text=white] at (axis cs:86.25, \yAnonymousOnly) {{27.7\%}};

    % Non-Anonymous bar
    \pgfmathsetmacro{\yNonAnonymousBottom}{\yNonAnonymous-0.125}
    \pgfmathsetmacro{\yNonAnonymousTop}{\yNonAnonymous+0.125}
    \draw[fill=blue!90, draw=white, line width=0.5pt] (axis cs:0, \yNonAnonymousBottom) rectangle (axis cs:2.5, \yNonAnonymousTop);
    \node[font=\tiny, text=white] at (axis cs:1.25, \yNonAnonymous) {{2\%}};
    \draw[fill=cyan!70, draw=white, line width=0.5pt] (axis cs:2.5, \yNonAnonymousBottom) rectangle (axis cs:6.9, \yNonAnonymousTop);
    \node[font=\tiny, text=white] at (axis cs:4.7, \yNonAnonymous) {{4\%}};
    \draw[fill=gray!80, draw=white, line width=0.5pt] (axis cs:6.9, \yNonAnonymousBottom) rectangle (axis cs:17.4, \yNonAnonymousTop);
    \node[font=\tiny, text=white] at (axis cs:12.149999999999999, \yNonAnonymous) {{10.5\%}};
    \draw[fill=orange!70, draw=white, line width=0.5pt] (axis cs:17.4, \yNonAnonymousBottom) rectangle (axis cs:62.0, \yNonAnonymousTop);
    \node[font=\small, text=white] at (axis cs:39.7, \yNonAnonymous) {{44.6\%}};
    \draw[fill=red!90, draw=white, line width=0.5pt] (axis cs:62.0, \yNonAnonymousBottom) rectangle (axis cs:100.0, \yNonAnonymousTop);
    \node[font=\small, text=white] at (axis cs:81.0, \yNonAnonymous) {{38.0\%}};
    \addlegendimage{area legend, fill=blue!90, draw=white, line width=0.5pt}
    \addlegendentry{Strongly disagree}
    \addlegendimage{area legend, fill=cyan!70, draw=white, line width=0.5pt}
    \addlegendentry{Disagree}
    \addlegendimage{area legend, fill=gray!80, draw=white, line width=0.5pt}
    \addlegendentry{Neither agree nor disagree}
    \addlegendimage{area legend, fill=orange!70, draw=white, line width=0.5pt}
    \addlegendentry{Agree}
    \addlegendimage{area legend, fill=red!90, draw=white, line width=0.5pt}
    \addlegendentry{Strongly agree}

    % Add sample size labels
    \node[font=\small, anchor=west] at (axis cs:100.5, \yNotComfortable) {(n=760)};
    \node[font=\small, anchor=west] at (axis cs:100.5, \yAnonymousOnly) {(n=1043)};
    \node[font=\small, anchor=west] at (axis cs:100.5, \yNonAnonymous) {(n=569)};
    \end{axis}
    \end{tikzpicture}
\caption{Agreement with the statement ``\emph{When viewing papers for which I am neither an author nor a reviewer, I sometimes find publicly-available, official reviews or comments useful},'' broken down by comfort commenting. (n=2372)}
\label{fig:q3q2_crosstab_combined}
\end{subfigure}
\caption{Comfort commenting on papers and agreement with comment usefulness.}
\label{fig:question3_crosstab_combined}
\end{figure}
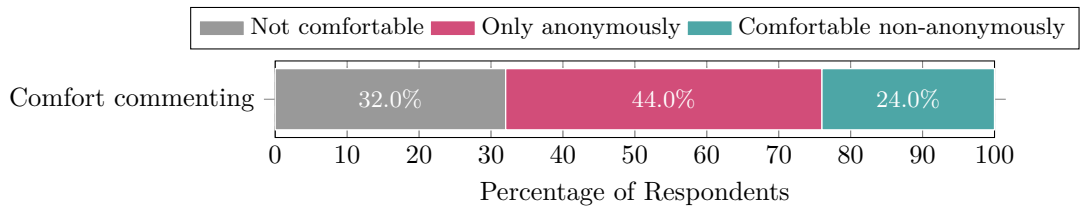
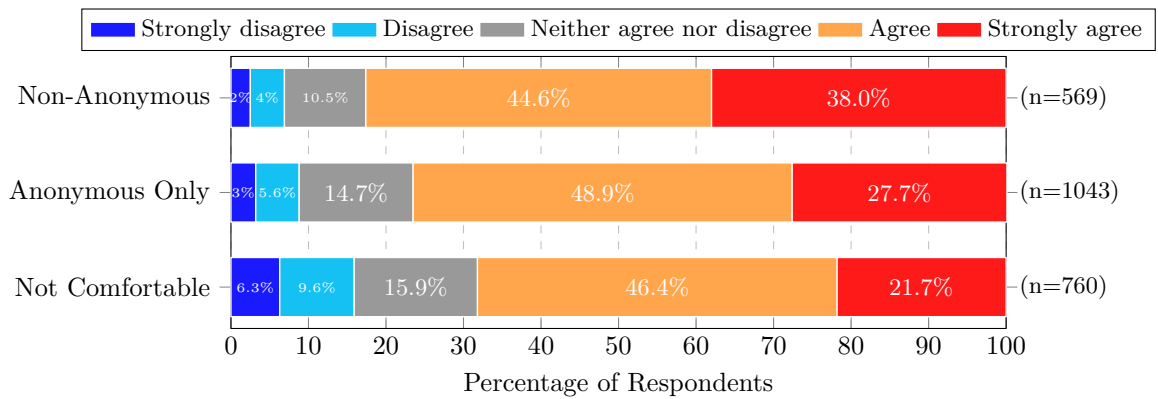

We asked ``\emph{In review settings where public commenting is allowed, how comfortable are you commenting on papers for which you are neither an author nor a reviewer?}'' The results are shown in \Cref{fig:question3_combined}. $32\%$ of respondents are not comfortable commenting, $44\%$ comment only anonymously, and $24\%$ are comfortable commenting non-anonymously. We compare these results to those of \citet{soergel2013open}. Comments could only be made non-anonymously in ICLR $2013$. Under those conditions, $45\%$ of respondents reported commenting happily, $40\%$ withheld comments entirely, and $15\%$ commented despite discomfort. Our result suggests that conference participants have grown more wary of commenting non-anonymously since $2013$ (perhaps due to the availability of anonymous commenting in recent iterations of ICLR).

\Cref{fig:q3q2_crosstab_combined} shows there is a  correlation between the willingness of respondents to leave comments and their experiences of having read useful comments in the past.

% Combined Question 3 and Q3 x Q2 Cross-tabulation Plot
% Note: This requires the subcaption package - add \usepackage{subcaption} to your LaTeX preamble

There were $30$ total responses about public commenting in the free-response question. Some mention the noise introduced in the process by anonymous commenters ($n=12$). A few mention that they rarely see public comments on their or others' papers anyway ($n=6$). Of those who feel negatively about public comments, they typically mention the possibility for anonymity to be misused -- authors can comment on their own papers to artificially praise them, and rivals can comment on papers to try to have them rejected. When commenting is allowed before the conference makes acceptance decisions, these sources of noise can impact the decisions. Two responses note that there is no clear method to raise issues with a conference paper once it has been accepted, aside from (anonymous) public commenting on OpenReview.

\subsection{Are respondents afraid of de-anonymization when reviewing?}
\label{subsec:wererespondentsafraid}

We ask respondents to rate their agreement with the statement ``\emph{When a conference might make my review public, I am afraid that I can be de-anonymized as a reviewer}.''
The results are shown in row (iv) of \Cref{fig:combined_likert}. We see that approximately $45\%$ of respondents disagree, and approximately $35\%$ agree.  
People consider anonymity to be quite important: \citet{soergel2013open} found that $84\%$ of respondents replied ``\emph{Yes}'' to the question ``\emph{It was important to me that my reviews be anonymous}.'' Given that most reviewers find it important that their reviews remain anonymous, $35\%$ fearing de-anonymization is quite relevant. 

We also break responses to this question down by career stage in \Cref{fig:career_stage_breakdown_7}. There does not appear to be a relationship between seniority and fear of de-anonymization.

In \Cref{fig:q3q7_crosstab}, we also break down responses to this question based on respondents' willingness to comment. We see that those who are comfortable commenting non-anonymously are also much less concerned with de-anonymization.

\careerStageBreakdown{Agreement with the statement ``\emph{When a conference might make my review public, I am afraid that I can be de-anonymized as a reviewer},'' broken down by career stage. ($n=1799$)}{fig:career_stage_breakdown_7}{706,236,507,350,12.3,32.0,21.7,26.2,7.8,16.9,34.3,19.9,23.3,5.5,11.4,32.0,21.1,26.0,9.5,10.6,36.3,23.1,23.1,6.9}{0.3}

\commentingFearCrossTab{Agreement with the statement ``\emph{When a conference might make my review public, I am afraid that I
can be de-anonymized as a reviewer},'' broken down by willingness to comment. ($n=2039$)}{fig:q3q7_crosstab}{10.8,34.4,17.5,27.5,9.8,640,9.0,28.9,24.4,28.4,9.2,919,20.2,37.9,22.3,16.2,3.3,480,249,668,443,515,164,2039}{0.2}

\subsection{How does open reviewing impact the quality of reviews?}
\label{subsec:howdoesopenreviewingimpact}

We ask respondents to rate their agreement with the statement ``\emph{When a conference might make my review public, I feel more incentivized to write a high quality review}.'' The results are shown in row (v) of \Cref{fig:combined_likert}. Overall, approximately $50\%$ of respondents say that open-reviewing incentivizes them to produce higher quality reviews, relative to approximately $15\%$ who disagree. \citet{soergel2013open} found that $45\%$ of reviewers said they felt more pressure to be clear and constructive in an open reviewing model, and the other $55\%$ said they felt the same level of pressure to be clear and constructive. 

We hypothesized that reviewers who are most fearful of de-anonymization feel the strongest incentives to improve their review quality, but we do not find a clear relationship between answers to these questions. 

% Combined Question 8 and Question 9 Topics Plot
% Note: This requires the subcaption package - add \usepackage{subcaption} to your LaTeX preamble
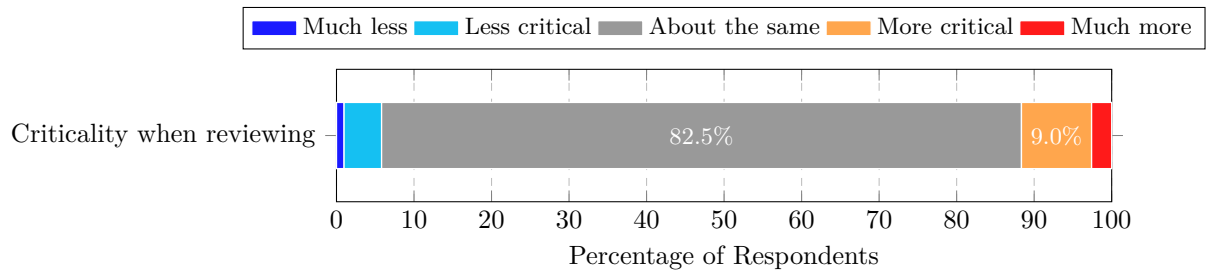
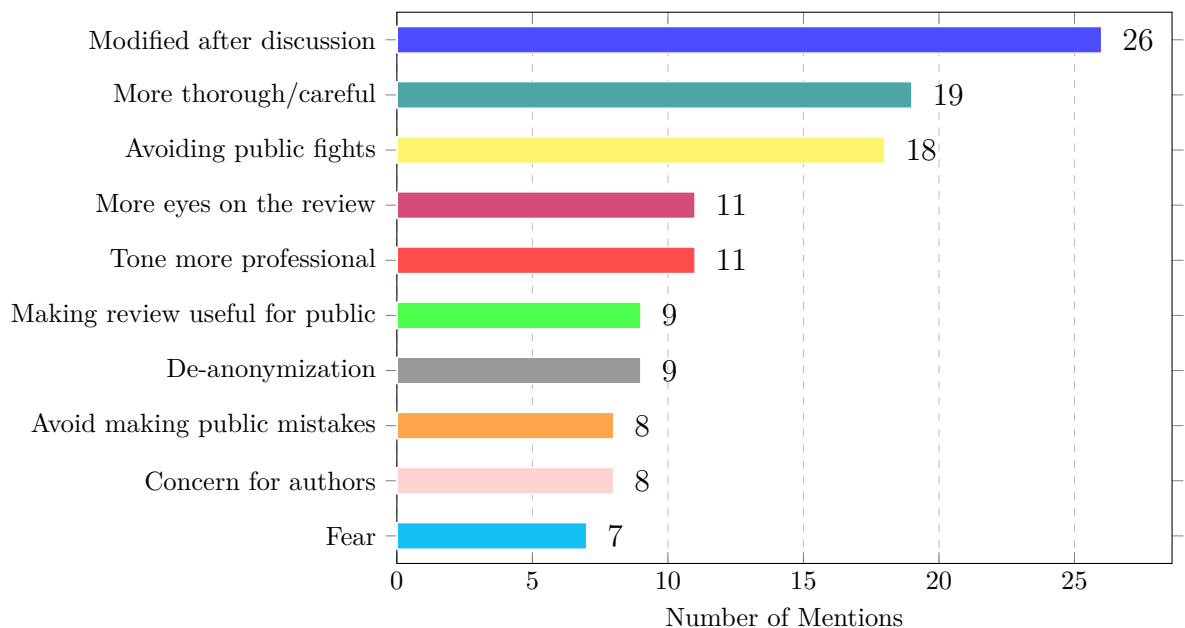
\begin{figure}
\centering
\begin{subfigure}[t]{0.9\textwidth}
\centering
\vspace{0.2cm}
    \begin{tikzpicture}
    \begin{axis}[
        xbar,
        xmin=0,
        xmax=100,
        width=0.8\textwidth,
        height=0.225\textwidth,
        xlabel={Percentage of Respondents},
        xlabel style={font=\normalsize},
        yticklabels={Criticality when reviewing},
        ytick={0},
        ymin=-0.25,
        ymax=0.25,
        bar width=6pt,
        bar shift=0pt,
        xmajorgrids=true,
        xminorgrids=true,
        grid style={dashed},
        clip=false,
        legend style={at={(0.5,1.15)}, anchor=south, legend columns=5, font=\small}
    ]
    
    % Define bar dimensions
    \def\barheight{0.125}
    \pgfmathsetmacro{\yBar}{0}
    \pgfmathsetmacro{\yBarBottom}{\yBar-\barheight}
    \pgfmathsetmacro{\yBarTop}{\yBar+\barheight}
    
    % Create stacked bar segments
    \draw[fill=blue!90, draw=white, line width=0.5pt] (axis cs:0, \yBarBottom) rectangle (axis cs:0.9779951100244498, \yBarTop);
    \draw[fill=cyan!70, draw=white, line width=0.5pt] (axis cs:0.9779951100244498, \yBarBottom) rectangle (axis cs:5.819070904645477, \yBarTop);
    \draw[fill=gray!80, draw=white, line width=0.5pt] (axis cs:5.819070904645477, \yBarBottom) rectangle (axis cs:88.36185819070904, \yBarTop);
    \node[font=\small, text=white] at (axis cs:47.09046454767726, \yBar) {{82.5\%}};
    \draw[fill=orange!70, draw=white, line width=0.5pt] (axis cs:88.36185819070904, \yBarBottom) rectangle (axis cs:97.4083129584352, \yBarTop);
    \node[font=\small, text=white] at (axis cs:92.88508557457212, \yBar) {{9.0\%}};
    \draw[fill=red!90, draw=white, line width=0.5pt] (axis cs:97.4083129584352, \yBarBottom) rectangle (axis cs:100.0, \yBarTop);
    \addlegendimage{area legend, fill=blue!90, draw=white, line width=0.5pt}
    \addlegendentry{Much less}
    \addlegendimage{area legend, fill=cyan!70, draw=white, line width=0.5pt}
    \addlegendentry{Less critical}
    \addlegendimage{area legend, fill=gray!80, draw=white, line width=0.5pt}
    \addlegendentry{About the same}
    \addlegendimage{area legend, fill=orange!70, draw=white, line width=0.5pt}
    \addlegendentry{More critical}
    \addlegendimage{area legend, fill=red!90, draw=white, line width=0.5pt}
    \addlegendentry{Much more}
    \end{axis}
    \end{tikzpicture}
\caption{Distribution of responses to ``\emph{Compared to similar conferences with closed reviewing policies, when I review at a conference with partially or fully open reviewing policies I am...}.'' ($n=2045$)}
\label{fig:question8_combined}
\vspace{0.3cm}
\end{subfigure}

\vspace{0.5cm}

\begin{subfigure}[t]{0.9\textwidth}
\centering
    \begin{tikzpicture}
    \begin{axis}[
        xbar,
        xmin=0,
        xmax=28.6,
        width=0.8\textwidth,
        height=0.6\textwidth,
        xlabel={Number of Mentions},
        xlabel style={font=\normalsize},
        yticklabels={Fear,Concern for authors,Avoid making public mistakes,De-anonymization,Making review useful for public,Tone more professional,More eyes on the review,Avoiding public fights,More thorough/careful,Modified after discussion},
        ytick={0,1,2,3,4,5,6,7,8,9},
        ymin=-0.5,
        ymax=9.5,
        bar width=10pt,
        bar shift=0pt,
        xmajorgrids=true,
        xminorgrids=true,
        grid style={dashed},
        clip=false
    ]
    
    % Create bars with different colors
    \addplot[fill=blue!70, draw=white, line width=0.5pt] coordinates {
        (0, 9) (26, 9)
    };
    \addplot[fill=teal!70, draw=white, line width=0.5pt] coordinates {
        (0, 8) (19, 8)
    };
    \addplot[fill=yellow!70, draw=white, line width=0.5pt] coordinates {
        (0, 7) (18, 7)
    };
    \addplot[fill=purple!70, draw=white, line width=0.5pt] coordinates {
        (0, 6) (11, 6)
    };
    \addplot[fill=red!70, draw=white, line width=0.5pt] coordinates {
        (0, 5) (11, 5)
    };
    \addplot[fill=green!70, draw=white, line width=0.5pt] coordinates {
        (0, 4) (9, 4)
    };
    \addplot[fill=gray!80, draw=white, line width=0.5pt] coordinates {
        (0, 3) (9, 3)
    };
    \addplot[fill=orange!70, draw=white, line width=0.5pt] coordinates {
        (0, 2) (8, 2)
    };
    \addplot[fill=pink!70, draw=white, line width=0.5pt] coordinates {
        (0, 1) (8, 1)
    };
    \addplot[fill=cyan!70, draw=white, line width=0.5pt] coordinates {
        (0, 0) (7, 0)
    };
    \node[font=\large, text=black, anchor=west] at (axis cs:26.429, 9) {26};
    \node[font=\large, text=black, anchor=west] at (axis cs:19.429, 8) {19};
    \node[font=\large, text=black, anchor=west] at (axis cs:18.429, 7) {18};
    \node[font=\large, text=black, anchor=west] at (axis cs:11.429, 6) {11};
    \node[font=\large, text=black, anchor=west] at (axis cs:11.429, 5) {11};
    \node[font=\large, text=black, anchor=west] at (axis cs:9.429, 4) {9};
    \node[font=\large, text=black, anchor=west] at (axis cs:9.429, 3) {9};
    \node[font=\large, text=black, anchor=west] at (axis cs:8.429, 2) {8};
    \node[font=\large, text=black, anchor=west] at (axis cs:8.429, 1) {8};
    \node[font=\large, text=black, anchor=west] at (axis cs:7.429, 0) {7};
    \end{axis}
    \end{tikzpicture}
\caption{ Top $10$ topics in response to ``\emph{If you have modified your level of criticism when reviewing at conferences with partially or fully open reviewing policies, please explain the reason(s) why}.'' ($n=162$)}
\label{fig:question9_topics_combined}
\end{subfigure}
\caption{Criticality when reviewing at open conferences and reasons mentioned.}
\label{fig:question8_9_combined}
\end{figure}

\questionEightBreakdown{Responses to ``\emph{Compared to similar conferences with closed reviewing policies, when I review at a conference with partially or fully open reviewing policies I am...},'' broken down by agreement with the statement ``\emph{When a conference might make my review public, I am afraid that I can be de-anonymized as a reviewer}.'' ($n=2044$)}{fig:question8_agreement_breakdown}{9.1,16.5,57.3,8.5,8.5,164,0.4,7.0,76.3,13.6,2.7,515,0.0,3.2,84.9,10.4,1.6,444,0.1,2.8,88.8,7.0,1.2,670,0.8,1.2,91.2,2.8,4.0,251}

We ask respondents to complete the sentence ``\emph{Compared to similar conferences with closed reviewing policies, when I review at a conference with partially or fully open reviewing policies I am...}.'' We show the results in \Cref{fig:question8_combined}. We see that approximately $6\%$ of respondents report being ``\emph{less critical of papers I review}'' or ``\emph{much less critical of papers I review},'' and double ($12\%$) report being ``\emph{more critical of papers I review}'' or ``\emph{much more critical of papers I review}.'' Most ($80\%$) do not modify their reviews to be more or less critical.

We also ask the free-response question ``\emph{If you have modified your level of criticism when reviewing at conferences with partially or fully open reviewing policies, please explain the reason(s) why}.'' We manually group the responses into common themes, and discard all responses which indicate only the ``None'' theme (e.g., responses like ``\emph{N/A}'' or ``\emph{I have not modified my level of criticism}''). \Cref{fig:question9_topics_combined} shows the top ten most common themes. Perhaps the most interesting in the top ten themes are ``avoiding public fights,'' ``making review useful for public,'' ``concern for authors,'' and ``de-anonymization.'' Many respondents to this question cite the potential for authors or public commenters to fight back against reviews perceived as being incorrect, and modify their level of criticism (typically becoming less critical) in order to avoid these fights. Others cite their desire to make the review useful for a wider range of people in the public, hence either increasing the level of criticism to be more thorough or decreasing criticisms that may offend others. The ``concern for authors'' topic is closely related to fears about resubmission bias; some reviewers tone down their criticisms when they believe the criticism may hurt the authors' ability to resubmit the paper elsewhere. Finally, some respondents are explicitly concerned with being de-anonymized, and fear retaliation or loss of reputation. 

Many of the free-response comments at the end of the survey (\Cref{fig:top_free_responses}) also touch on the impact of open reviewing policies on review quality. Some advance the notion that reviewers are not incentivized by the public nature of the anonymized reviews. A large number of comments ($n=47$) indicate that open reviewing policies are not the most important factor driving review quality. Many comments recommend instead that stronger incentives are now needed to encourage reviewers to exert a proper amount of effort, with $40$ mentioning the need for increased accountability to reviewers and $31$ mentioning introducing more direct incentives. A surprising number of respondents ($n=61$) suggest de-anonymizing all reviewers, or at least reviewers deemed to be most in violation of reviewing policies.

We also report the distribution of answers to ``\emph{Compared to similar conferences with closed reviewing policies, when I review at a conference with partially or fully open reviewing policies I am...},'' broken down by fear of de-anonymization. This is shown in \Cref{fig:question8_agreement_breakdown}. Those who are most fearful of de-anonymization are the most likely to report modifying their level of criticism (either more or less). Among those who strongly agree that they are afraid of being de-anonymized, approximately $26\%$ report lowering their level of criticism in open review, with approximately $17\%$ reporting a higher level of criticism. In all other cases, reviewers report nearly doubling their level of criticism. Those who are strongly fearful of being de-anonymized are most likely to modify their behavior, and are more likely to reduce criticism rather than increase it. The remainder of reviewers are somewhat \emph{more} critical. The overall effect is that more reviewers report increasing their level of criticism.

\subsection{Does open reviewing result in a more fair decision process?}
\label{subsec:doesopenreviewingresultin}

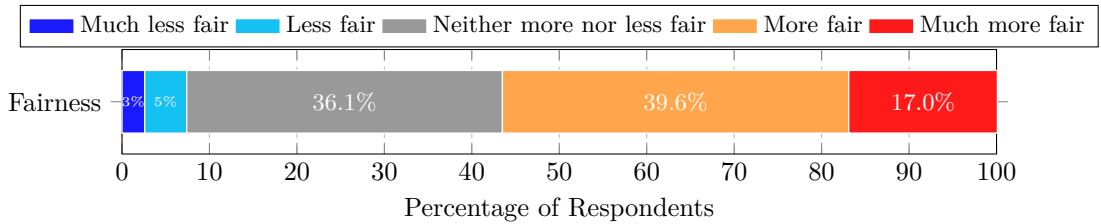
\singleLikertBar{Distribution of responses to: ``\emph{Compared to closed reviewing policies, partially and fully open reviewing policies result in a decision process that is....}'' ($n=2376$)}{fig:question_4_choice}{Much less fair,Less fair,Neither more nor less fair,More fair,Much more fair}{2.6,4.8,36.1,39.6,17.0}{Fairness}{likert}{2.5,22.0,35.0,64.0,79.0}

We ask the respondents to indicate their opinions on fairness of the decision process by completing the statement ``\emph{Compared to closed reviewing policies, partially and fully open reviewing policies result in a decision process that is....}'' These results are shown in \Cref{fig:question_4_choice}.
We find that approximately $57\%$ of respondents say that the process is more fair, while only approximately $8\%$ say the process is ``less fair [or] much less fair.''  \citet{soergel2013open} found that approximately $75\%$ of respondents thought the open reviewing process was overall more fair, and approximately $25\%$ said the process was neither more nor less fair.

In the final free response, there are $20$ comments mentioning something about fairness of the process. Among the ones that support the fairness of open reviewing policies, the ones that explain a rationale typically equate fairness and transparency. However, some comments cite a possible source of unfairness: the possibility of unfair, low-effort, or fraudulent reviews being attached to papers in perpetuity. Five comments cite the outsized impact this effect can have on researchers that have less prestige (e.g., junior researchers or non-male researchers).

\subsection{Does open reviewing compel extra experiments?}
\label{subsec:doesopenreviewingcompel}

We ask respondents to rate their agreement with the statement ``\emph{Compared to a closed reviewing model, I feel that open reviewing models compel me to complete additional, unnecessary analyses or experiments during the discussion phase on papers I have submitted}.''
Row (vi) of \Cref{fig:combined_likert} shows the results. We find approximately $33\%$ of respondents agree with the statement, while approximately $29\%$ disagree.  
We also consider the possibility that agreement with this statement depends on career stage, but find no clear relationship.

In the free text responses (\Cref{fig:top_free_responses}), $36$ respondents mention that the design of ICLR's discussion phase results in increased stress or requires additional work relative to other conferences. These respondents generally do not feel that this extra work comes from the \emph{openness} of ICLR's process, but rather from the asynchronous, multi-step discussion policies. One respondent mentions stress from public comments requesting large numbers of experiments, but in general the primary complaint voiced about the discussion phase is to limit the number of synchronous rounds instead of allowing unlimited replies. 

\subsection{Do participants refrain from submitting to fully open review, and why?}
\label{subsec:doparticipantsrefrain}

\begin{figure}
\centering
\begin{subfigure}[t]{0.9\textwidth}
\centering
\vspace{0.2cm}
    \begin{tikzpicture}
    \begin{axis}[
        xbar,
        xmin=0,
        xmax=100,
        width=0.8\textwidth,
        height=0.2\textwidth,
        xlabel={Percentage of Respondents},
        xlabel style={font=\normalsize},
        yticklabels={Refrain from open conferences},
        ytick={0},
        ymin=-0.25,
        ymax=0.25,
        bar width=6pt,
        bar shift=0pt,
        xmajorgrids=true,
        xminorgrids=true,
        grid style={dashed},
        clip=false,
        legend style={at={(0.5,1.15)}, anchor=south, legend columns=2, font=\small}
    ]
    
    % Define bar dimensions (taller bar)
    \def\barheight{0.125}
    \pgfmathsetmacro{\yBar}{0}
    \pgfmathsetmacro{\yBarBottom}{\yBar-\barheight}
    \pgfmathsetmacro{\yBarTop}{\yBar+\barheight}
    
    % Create stacked bar segments
    \draw[fill=gray!80, draw=white, line width=0.5pt] (axis cs:0, \yBarBottom) rectangle (axis cs:71.6, \yBarTop);
    \draw[fill=purple!70, draw=white, line width=0.5pt] (axis cs:71.6, \yBarBottom) rectangle (axis cs:100.0, \yBarTop);
    
    % Add percentage labels on the bars (always show both)
    \node[font=\small, text=white] at (axis cs:35.8, \yBar) {{71.6\%}};
    \node[font=\small, text=white] at (axis cs:85.8, \yBar) {{28.4\%}};
    
    % Add legend
    \addlegendimage{area legend, fill=gray!80, draw=white, line width=0.5pt}
    \addlegendentry{No}
    \addlegendimage{area legend, fill=purple!70, draw=white, line width=0.5pt}
    \addlegendentry{Yes}
    
    \end{axis}
    \end{tikzpicture}
\caption{Distribution of responses to ``\emph{Do you ever refrain from submitting to fully open conferences that release rejected submissions and their reviews?}'' (n=2187)}
\label{fig:question14_combined}
\vspace{0.3cm}
\end{subfigure}

\vspace{0.5cm}

\begin{subfigure}[t]{0.9\textwidth}
\centering
    \begin{tikzpicture}
    \begin{axis}[
        xbar,
        xmin=0,
        xmax=258.5,
        width=0.8\textwidth,
        height=0.55\textwidth,
        xlabel={Number of Mentions},
        xlabel style={font=\normalsize},
        yticklabels={Paper may be erroneous,Embarrassment,Submitting to get feedback,Biases future readers,Avoiding public rejection,Scooping,Reputation,Increased personal threshold,Reviews are unfair,Resubmission bias},
        ytick={0,1,2,3,4,5,6,7,8,9},
        ymin=-0.5,
        ymax=9.5,
        bar width=10pt,
        bar shift=0pt,
        xmajorgrids=true,
        xminorgrids=true,
        grid style={dashed},
        clip=false
    ]
    
    % Create bars with different colors
    \addplot[fill=blue!70, draw=white, line width=0.5pt] coordinates {
        (0, 0) (15, 0)
    };
    \addplot[fill=teal!70, draw=white, line width=0.5pt] coordinates {
        (0, 1) (16, 1)
    };
    \addplot[fill=yellow!70, draw=white, line width=0.5pt] coordinates {
        (0, 2) (17, 2)
    };
    \addplot[fill=purple!70, draw=white, line width=0.5pt] coordinates {
        (0, 3) (17, 3)
    };
    \addplot[fill=red!70, draw=white, line width=0.5pt] coordinates {
        (0, 4) (19, 4)
    };
    \addplot[fill=green!70, draw=white, line width=0.5pt] coordinates {
        (0, 5) (21, 5)
    };
    \addplot[fill=gray!80, draw=white, line width=0.5pt] coordinates {
        (0, 6) (26, 6)
    };
    \addplot[fill=orange!70, draw=white, line width=0.5pt] coordinates {
        (0, 7) (35, 7)
    };
    \addplot[fill=pink!70, draw=white, line width=0.5pt] coordinates {
        (0, 8) (72, 8)
    };
    \addplot[fill=cyan!70, draw=white, line width=0.5pt] coordinates {
        (0, 9) (235, 9)
    };
    \node[font=\large, text=black, anchor=west] at (axis cs:17, 0) {15};
    \node[font=\large, text=black, anchor=west] at (axis cs:18, 1) {16};
    \node[font=\large, text=black, anchor=west] at (axis cs:19, 2) {17};
    \node[font=\large, text=black, anchor=west] at (axis cs:19, 3) {17};
    \node[font=\large, text=black, anchor=west] at (axis cs:21, 4) {19};
    \node[font=\large, text=black, anchor=west] at (axis cs:23, 5) {21};
    \node[font=\large, text=black, anchor=west] at (axis cs:28, 6) {26};
    \node[font=\large, text=black, anchor=west] at (axis cs:37, 7) {35};
    \node[font=\large, text=black, anchor=west] at (axis cs:74, 8) {72};
    \node[font=\large, text=black, anchor=west] at (axis cs:237, 9) {235};
    \end{axis}
    \end{tikzpicture}
\caption{Top $10$ topics in response to ``\emph{If you ever refrain from submitting to fully open conferences that release rejected submissions and their reviews, please explain the reason(s) why.}'' (n=411)}
\label{fig:question15_topics_combined}
\end{subfigure}
\caption{Refraining from submission to fully open conferences and reasons mentioned.}
\label{fig:question14_15_combined}
\end{figure}
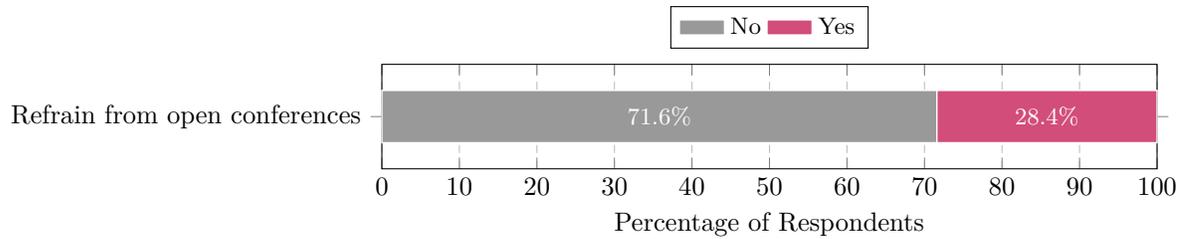
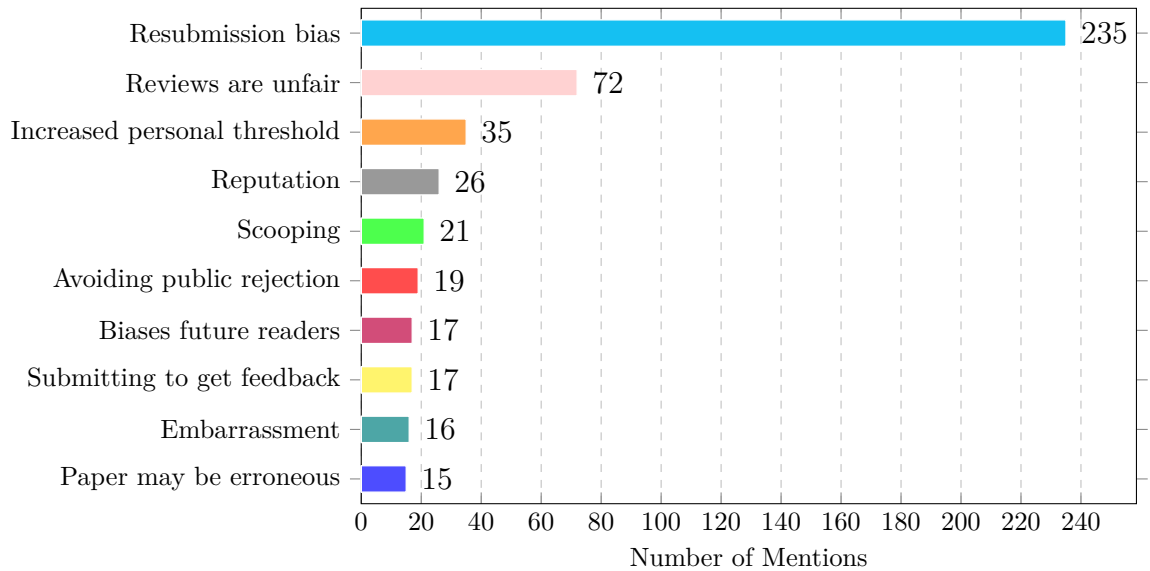

We ask authors ``\emph{Do you ever refrain from submitting to fully open conferences that release rejected submissions and their reviews?}''
As seen in \Cref{fig:question14_combined}, we find that $28.4\%$ of authors report refraining from submitting to fully open conferences such as ICLR. 

We ask a follow-up, free-response question: ``\emph{If you ever refrain from submitting to fully open conferences that release rejected submissions and their reviews, please explain the reason(s) why.}'' We manually group the responses into common themes, and discard all responses which indicate only the ``None'' theme (e.g., responses like ``\emph{N/A}'' or ``\emph{I have not refrained}'').
\Cref{fig:question15_topics_combined} shows the top $10$ topics in the responses. More than half of those who refrained mention the possibility of \emph{resubmission bias}, or the well-documented effect that when reviewers in a later conference know a paper has already been rejected, their evaluations are  negatively biased on average \citep{stelmakh2021prior}. Not only are many participants afraid of resubmission bias, many indicate they have been victims of this phenomenon. Some respondents report experiencing an extreme version of resubmission bias, in which they or a close collaborator substantially modified a resubmitted paper but still received verbatim copies of prior reviews from a fully open conference. 

The second most-popular rationale considers that negative comments on rejected papers will remain in the public record for that paper, even if the comments are not true or if the paper has been modified since. In addition, participants mention that the reviewing process can be very random, and can include reviewers with conflicts of interest who deliberately sabotage a paper. Such comments remain attached to a rejected submission forever. This concern often co-occurs with concerns about resubmission bias; the problem is not just that correct rejections follow a paper, but that unfair or deliberately-targeted rejections can ruin a paper's chances of resubmission.

The third top response reports that many refrain from submitting papers that they feel are not fully complete. Some report seeing this as positive, and some report seeing this as negative. On the positive side, many comments mention the reduction in workload for reviewers. However, others report that they like to submit papers that they know are not high quality, in order to receive feedback from the community or to give junior co-authors experience with the peer-review system. The comments give an overall sense that fully open conferences can deter authors from submitting lower-quality manuscripts, for better or for worse. 

Many comments mention the idea that having a non-anonymized, rejected manuscript available on the internet can negatively impact a researcher's reputation, especially early in their career. Some of the other topics mentioned include: possibility of scooping, desire to avoid embarrassment, and negative mental health impacts. 

% Combined Question 14 and Question 15 Topics Plot
% Note: This requires the subcaption package - add \usepackage{subcaption} to your LaTeX preamble

\subsection{How does AI interact with open reviewing?}
\label{subsec:howdollms}

We ask two questions about reducing irresponsible AI usage; one asks about AI usage by authors, and one asks about AI usage by reviewers. 

\begin{figure}[H]
\centering
\begin{subfigure}[t]{0.9\textwidth}
\centering
\vspace{0.2cm}
    \begin{tikzpicture}
    \begin{axis}[
        xbar,
        xmin=0,
        xmax=100,
        width=0.8\textwidth,
        height=0.2\textwidth,
        xlabel={Percentage of Respondents},
        xlabel style={font=\normalsize},
        yticklabels={Deter/encourage},
        ytick={0},
        ymin=-0.15,
        ymax=0.15,
        bar width=6pt,
        bar shift=0pt,
        xmajorgrids=true,
        xminorgrids=true,
        grid style={dashed},
        clip=false,
        legend style={at={(0.5,1.02)}, anchor=south, legend columns=5, font=\small}
    ]
    
    % Define bar dimensions
    \def\barheight{0.08}
    \pgfmathsetmacro{\yBar}{0}
    \pgfmathsetmacro{\yBarBottom}{\yBar-\barheight}
    \pgfmathsetmacro{\yBarTop}{\yBar+\barheight}
    
    % Create stacked bar segments
    \draw[fill=blue!90, draw=white, line width=0.5pt] (axis cs:0, \yBarBottom) rectangle (axis cs:7.176023638666104, \yBarTop);
    \pgfmathparse{7.176023638666104>=13}
    \ifnum\pgfmathresult=1
        \node[font=\small, text=white] at (axis cs:3.588011819333052, \yBar) {\formatPctLabel{7.2}\%};
    \else
        \pgfmathparse{7.176023638666104>0}
        \ifnum\pgfmathresult=1 \node[font=\tiny, text=white] at (axis cs:3.588011819333052, \yBar) {\formatPctLabel{7.2}\%}; \fi
    \fi
    \draw[fill=cyan!70, draw=white, line width=0.5pt] (axis cs:7.176023638666104, \yBarBottom) rectangle (axis cs:38.9193752638244, \yBarTop);
    \pgfmathparse{31.743351625158294>=13}
    \ifnum\pgfmathresult=1
        \node[font=\small, text=white] at (axis cs:23.04769945124525, \yBar) {\formatPctLabel{31.7}\%};
    \else
        \pgfmathparse{31.743351625158294>0}
        \ifnum\pgfmathresult=1 \node[font=\tiny, text=white] at (axis cs:23.04769945124525, \yBar) {\formatPctLabel{31.7}\%}; \fi
    \fi
    \draw[fill=gray!80, draw=white, line width=0.5pt] (axis cs:38.9193752638244, \yBarBottom) rectangle (axis cs:89.0671169269734, \yBarTop);
    \pgfmathparse{50.147741663149006>=13}
    \ifnum\pgfmathresult=1
        \node[font=\small, text=white] at (axis cs:63.9932460953989, \yBar) {\formatPctLabel{50.1}\%};
    \else
        \pgfmathparse{50.147741663149006>0}
        \ifnum\pgfmathresult=1 \node[font=\tiny, text=white] at (axis cs:63.9932460953989, \yBar) {\formatPctLabel{50.1}\%}; \fi
    \fi
    \draw[fill=orange!70, draw=white, line width=0.5pt] (axis cs:89.0671169269734, \yBarBottom) rectangle (axis cs:97.21401435204727, \yBarTop);
    \pgfmathparse{8.14689742507387>=13}
    \ifnum\pgfmathresult=1
        \node[font=\small, text=white] at (axis cs:93.14056563951033, \yBar) {\formatPctLabel{8.1}\%};
    \else
        \pgfmathparse{8.14689742507387>0}
        \ifnum\pgfmathresult=1 \node[font=\tiny, text=white] at (axis cs:93.14056563951033, \yBar) {\formatPctLabel{8.1}\%}; \fi
    \fi
    \draw[fill=red!90, draw=white, line width=0.5pt] (axis cs:97.21401435204727, \yBarBottom) rectangle (axis cs:100.0, \yBarTop);
    \pgfmathparse{2.7859856479527227>=13}
    \ifnum\pgfmathresult=1
        \node[font=\small, text=white] at (axis cs:98.60700717602364, \yBar) {\formatPctLabel{2.8}\%};
    \else
        \pgfmathparse{2.7859856479527227>0}
        \ifnum\pgfmathresult=1 \node[font=\tiny, text=white] at (axis cs:98.60700717602364, \yBar) {\formatPctLabel{2.8}\%}; \fi
    \fi
    \addlegendimage{area legend, fill=blue!90, draw=white, line width=0.5pt}
    \addlegendentry{Strongly deter}
    \addlegendimage{area legend, fill=cyan!70, draw=white, line width=0.5pt}
    \addlegendentry{Deter}
    \addlegendimage{area legend, fill=gray!80, draw=white, line width=0.5pt}
    \addlegendentry{Neither deter nor encourage}
    \addlegendimage{area legend, fill=orange!70, draw=white, line width=0.5pt}
    \addlegendentry{Encourage}
    \addlegendimage{area legend, fill=red!90, draw=white, line width=0.5pt}
    \addlegendentry{Strongly encourage}
    \end{axis}
    \end{tikzpicture}
\caption{Responses to ``\emph{To what extent do partially or fully open reviewing policies deter—versus encourage—reviewers from submitting reviews generated entirely using large language models?} ($n=2369$)}
\label{fig:question5_likert_combined}
\vspace{0.3cm}
\end{subfigure}

\vspace{0.5cm}

\begin{subfigure}[t]{0.9\textwidth}
\centering
    \begin{tikzpicture}
    \begin{axis}[
        xbar,
        xmin=0,
        xmax=320.1,
        width=0.8\textwidth,
        height=0.4\textwidth,
        xlabel={Number of Respondents},
        xlabel style={font=\normalsize},
        yticklabels={Prefer to deter,Prefer to retain,Do not believe deters slop,Do not believe deters legitimate},
        ytick={3,2,1,0},
        ymin=-0.5,
        ymax=3.5,
        bar width=8pt,
        bar shift=0pt,
        xmajorgrids=true,
        xminorgrids=true,
        grid style={dashed},
        clip=false
    ]
    
    % Create horizontal bars (biggest on top: y=3, smallest on bottom: y=0)
    % Bar 1 at y=3
    \draw[fill=blue!90, draw=white, line width=0.5pt] (axis cs:0, 2.7) rectangle (axis cs:291, 3.3);
    \pgfmathparse{291>58.2}
    \ifnum\pgfmathresult=1
        \node[font=\small, text=white] at (axis cs:{291/2}, 3) {291 (44.9\%)};
    \else
        \node[font=\small, text=black, anchor=west] at (axis cs:291, 3) {291 (44.9\%)};
    \fi
    % Bar 2 at y=2
    \draw[fill=purple!70, draw=white, line width=0.5pt] (axis cs:0, 1.7) rectangle (axis cs:246, 2.3);
    \pgfmathparse{246>58.2}
    \ifnum\pgfmathresult=1
        \node[font=\small, text=white] at (axis cs:{246/2}, 2) {246 (38.0\%)};
    \else
        \node[font=\small, text=black, anchor=west] at (axis cs:246, 2) {246 (38.0\%)};
    \fi
    % Bar 3 at y=1
    \draw[fill=teal!70, draw=white, line width=0.5pt] (axis cs:0, 0.7) rectangle (axis cs:182, 1.3);
    \pgfmathparse{182>58.2}
    \ifnum\pgfmathresult=1
        \node[font=\small, text=white] at (axis cs:{182/2}, 1) {182 (28.1\%)};
    \else
        \node[font=\small, text=black, anchor=west] at (axis cs:182, 1) {182 (28.1\%)};
    \fi
    % Bar 4 at y=0
    \draw[fill=red!80, draw=white, line width=0.5pt] (axis cs:0, -0.3) rectangle (axis cs:169, 0.3);
    \pgfmathparse{169>58.2}
    \ifnum\pgfmathresult=1
        \node[font=\small, text=white] at (axis cs:{169/2}, 0) {169 (26.1\%)};
    \else
        \node[font=\small, text=black, anchor=west] at (axis cs:169, 0) {169 (26.1\%)};
    \fi
    \end{axis}
    \end{tikzpicture}
\caption{Responses to ``\emph{Publicly releasing rejected submissions with author names may deter authors from submitting AI slop or other types of clearly low-quality (e.g., incomplete) submissions, but it may also deter legitimate submissions. Considering this situation, which of the following statements reflect your views? Select all that apply.}'' ($n=648$)}
\label{fig:question6_multiselect_combined}
\end{subfigure}
\caption{Deterrence of low-quality, AI-generated review content and manuscript submissions.}
\label{fig:question5_6_combined}
\end{figure}
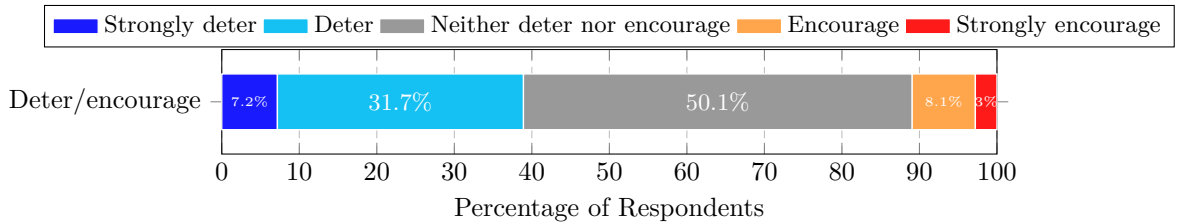
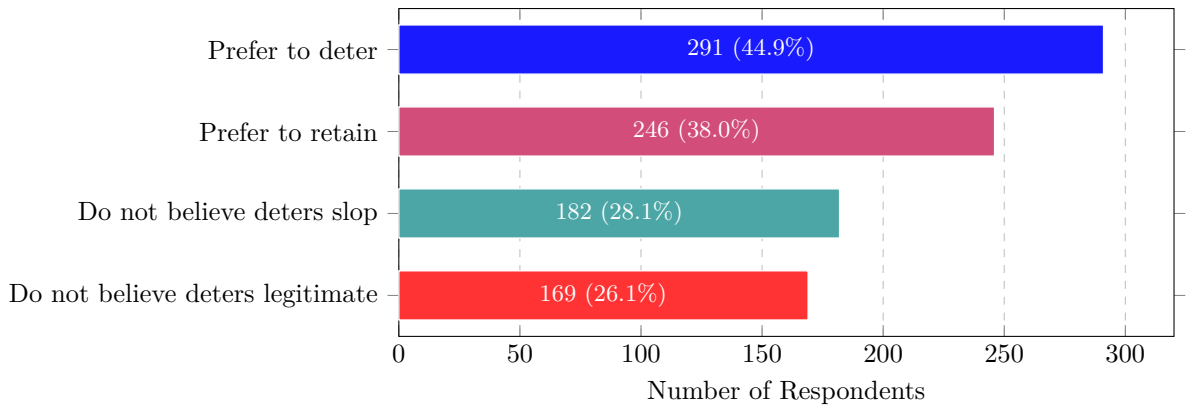

The first question asks ``\emph{To what extent do partially or fully open reviewing policies deter—versus encourage—reviewers from submitting reviews generated entirely using large language models?}'' The results can be seen in \Cref{fig:question5_likert_combined}.
We find that approximately $39\%$ of respondents believe that partially or fully open policies ``\emph{deter}'' or ``\emph{strongly deter}'' this behavior, and approximately $11\%$ say they encourage it.

% Combined Question 5 and Question 6 Plot

We also ask ``\emph{Publicly releasing rejected submissions with author names may deter authors from submitting AI slop or other types of clearly low-quality (e.g., incomplete) submissions, but it may also deter legitimate submissions. Considering this situation, which of the following statements reflect your views? Select all that apply.}'' The options provided are: ``\emph{I prefer that conferences deter AI slop and other extremely low-quality submissions, even if there is potential to lose some legitimate submissions},'' ``\emph{I prefer that conferences retain legitimate submissions, even if there is potential to receive more AI slop and other extremely low-quality submissions},'' ``\emph{I do not believe releasing rejected submissions with names deters AI slop and other extremely low-quality submissions},'' and ``\emph{I do not believe releasing rejected submissions with names deters legitimate submissions}.''
Although this question was added after the survey launched, $660$ people started and completed the survey after this question was added and $648$ of those respondents answered this question.
Results are shown in \Cref{fig:question6_multiselect_combined}. We see approximately $45\%$ of respondents prefer to deter AI slop and extremely low quality papers, even if some legitimate submissions are deterred, while $38\%$ prefer to retain legitimate submissions even at the expense of increased low quality submissions. A majority of respondents believe that fully open reviewing has some deterrence effect. Deterrence continues to beat retention even if we restrict to those who believe that releasing rejected papers with names has a deterrence effect on both legitimate and illegitimate submissions. Some of the  free responses ($n=9$) mention that AI slop papers can just be desk rejected, and $11$ mention that open reviewing cuts down on submissions that the authors know to be bad.

Altogether, the survey participants mostly believe that open reviewing policies can deter both legitimate and slop submissions, but weakly prefer to continue deterring slop even at the expense of some legitimate submissions. They also believe that open policies can help deter AI-generated reviews.

We also ask the question ``\emph{Should conferences provide official, LLM-generated reviews to supplement human expert reviews? If so, who should be able to see them? (Select all that apply)}.'' Large computer science conferences have begun experimenting with ways to use AI in their workflows \citep{thakkar2025LLMFeedbackICLR}, and one major conference provided clearly-labeled, AI-generated reviews directly alongside human reviews \citep{AAAI_2025_AI_review_assessment}. We aim to understand respondents' preferences about this practice. \Cref{fig:question20_multiselect} shows the responses. Slightly under half of respondents are supportive of this practice. The most-supported group to show LLM-generated reviews is meta-reviewers, followed by authors. The top $5$ most common response sets are ``\emph{Authors, General public, Meta-reviewers, Reviewers}'' ($n=267$ or $11.4\%$ of all respondents), 
``\emph{Authors, Meta-reviewers, Reviewers}'' ($n=170$ or $7.3\%$ of all respondents), ``\emph{General public}'' ($n=143$ or $6.1\%$ of all respondents), ``\emph{Meta-reviewers}'' ($n=97$ or $4.13\%$ of all respondents), and ``\emph{Authors, Meta-reviewers}'' ($n=82$ or $3.50\%$ of all respondents).
We also analyze the responses of area chairs and senior area chairs separately. This population has a very similar distribution to the overall respondent population, but they are slightly more supportive of official, AI-generated reviews overall ($3\%$ increase in support).

\multiSelectBarFive{Responses to ``\emph{Should conferences provide official, LLM-generated reviews to supplement human expert reviews? If so, who should be able to see them? (Select all that apply)}.'' ($n=2346$)}{fig:question20_multiselect}{Do not support LLM reviews,Meta-reviewers,Authors,Reviewers,General public}{56.0,31.3,27.8,25.2,20.3}{1313,734,653,592,477}

\subsection{How do respondents rank potential impacts of open peer review?}
\label{subsec:howdorespondentsrank}

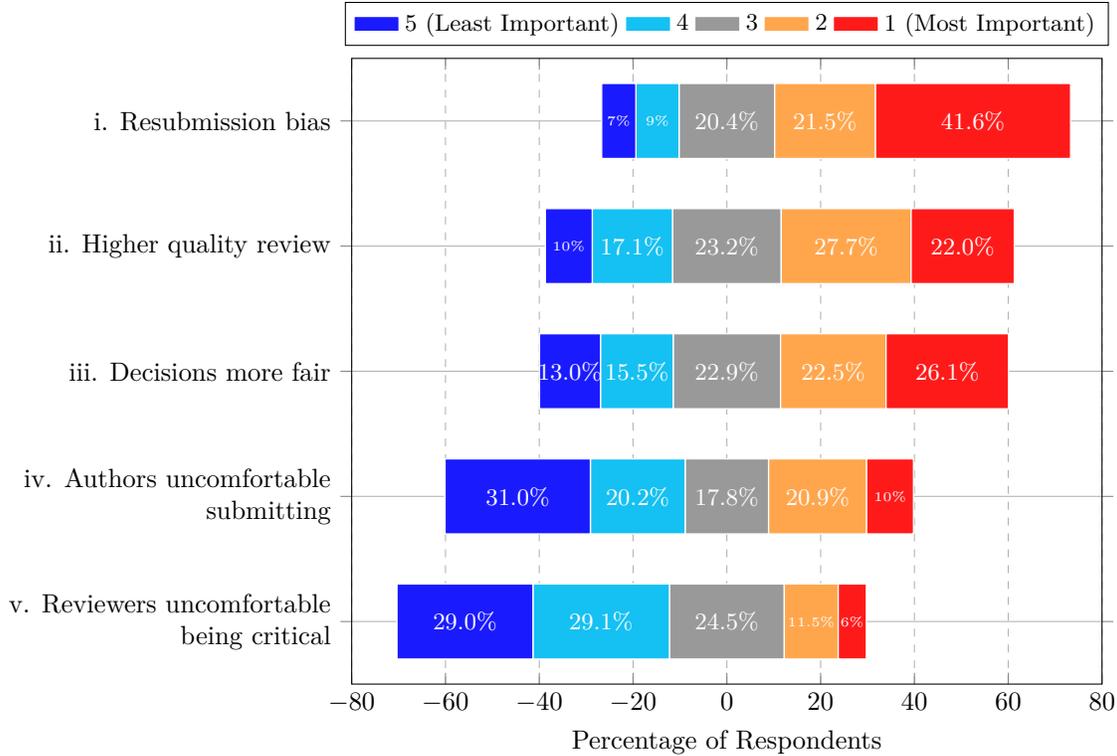
\questionSixteenRankingPlot{Ranks assigned to possible impacts of open peer-review policies. ($n=2299$)}{fig:question16_ranking}{i. Resubmission bias,ii. Higher quality review,iii. Decisions more fair,iv. Authors uncomfortable | submitting,v. Reviewers uncomfortable | being critical}{41.6,21.5,20.4,9.2,7.3,22.0,27.7,23.2,17.1,10.0,26.1,22.5,22.9,15.5,13.0,10.0,20.9,17.8,20.2,31.0,6.0,11.5,24.5,29.1,29.0}

We ask participants: ``\emph{Consider the potential impacts of fully open reviewing policies that release rejected submissions and reviews. Please rank these outcomes based on how much they matter to you overall—taking into account both how likely each one seems to you, and how good or bad it would be if it happened. Rank from 1 (matters most) to 5 (matters least). Use each rank exactly once.}'' The options given are: ``\emph{Authors might not feel comfortable submitting their work to the venue},'' ``\emph{Reviewers might not feel comfortable expressing critical opinions},'' ``\emph{Reviewers might have more incentive to write a high quality review},'' ``\emph{The decision-making process might be more fair},'' and ``\emph{Authors with rejected manuscripts might experience resubmission bias or duplicated reviews at future venues}.''
\Cref{fig:question16_ranking} shows the percentage of responses with each rank assigned to each possible impact.

Resubmission bias is the clearest concern for respondents, being ranked first by approximately $42\%$ of respondents and either first or second by approximately $63\%$.  
The positive impacts (incentives to write a high quality review and fairness of the decision-making process), rank in the top two for approximately $50\%$ of respondents. Authors being less comfortable submitting to the venue and reviewers' reduced willingness to express critical opinions rank fourth or fifth $50\%$ and $60\%$ of the time.  

As discussed earlier, we find that approximately $29\%$ of authors have refrained from submitting to fully open conferences, and only one-quarter of respondents say they do \emph{not} believe fully open policies deter legitimate submissions. If we restrict the ranking analysis to those participants who have refrained from submitting to a fully open conference, we find that the rankings are fairly consistent with the full population, but authors' comfort submitting to the venue is ranked second (behind resubmission bias).

This ranking activity suggests that mitigating resubmission bias is the number one priority, but aside from resubmission bias many conference participants weigh the positive aspects of open reviewing over the negatives. There is a large minority of conference participants that also feels that open reviewing conferences' disincentivization of submissions is more relevant than the possible positive outcomes. 

\subsection{What else do respondents say in the free-response section?}
\label{subsec:whatelse}

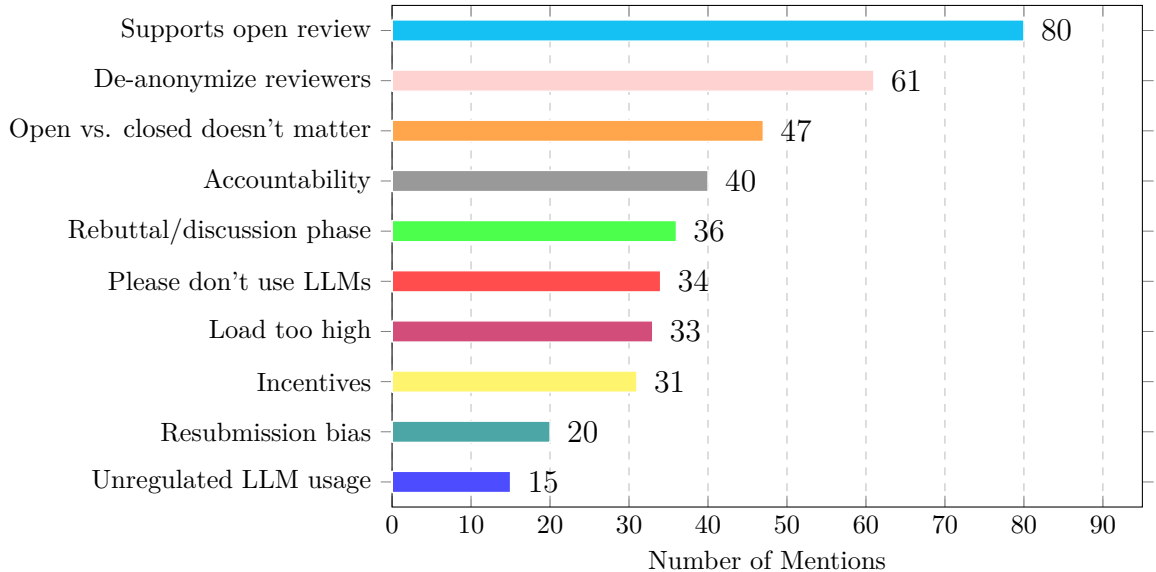
\topicDistributionBar{Unregulated LLM usage,Resubmission bias,Incentives,Load too high,Please don't use LLMs,Rebuttal/discussion phase,Accountability,Open vs. closed doesn't matter,De-anonymize reviewers,Supports open review}{15,20,31,33,34,36,40,47,61,80}{Number of Mentions}{Top $10$ topics mentioned in responses to ``\emph{Do you have any additional thoughts you would like to express about open or closed reviewing practices?}'' ($n=494$)}{fig:top_free_responses}{95}{1}

We ask the following  free-response question at the end of the survey: ``\emph{Do you have any additional thoughts you would like to express about open or closed reviewing practices?}''  We perform a qualitative coding of the responses, with results displayed in \Cref{fig:top_free_responses}. We received $494$ free text responses in this field. Although we have already discussed many of the themes surfaced in the free-response text, we highlight a few additional topics that are not directly related to the other themes in the survey.

The highest-frequency topic, with $80$ unique responses, is explicit, enthusiastic support for open reviewing policies. Many ($47$ respondents) mention that open vs. closed review is not the primary question, but that there are other, more pressing concerns they have with the peer-review process. Many of these comments also support improving the mechanisms for accountability of reviewers ($40$ responses), or better incentivizing reviewers or authors ($31$ responses). One common theme ($36$ responses) centers around suggested changes to the \emph{discussion period}. After all reviews are released, authors and reviewers participate in a two week discussion period. Both parties post unlimited responses to each other, and the authors may update their manuscript. Many of the free-response comments suggest limiting discussions to a single round between authors and reviewers, since multi-round discussion periods require a large amount of work for both parties. One other common theme in these responses is to suggest that conferences should work harder to ensure that reviewers respond to authors' comments and updates during the discussion phase.

\subsection{How would participants design a future conference?}
\label{subsec:howwouldparticipantsdesign}

Near the end of the survey, we have a section that begins with the sentence ``\emph{In this section, you will select the set of policies you would recommend to future conference organizers.}'' We then ask four questions: ``\emph{How would you set the visibility of submissions?}'', ``\emph{How would you set the visibility
of anonymized reviews?}'', and ``\emph{How would you set the policy on public commenting?}'' For each of these questions, we provide respondents with a choice describing the closed, partially open, or fully open option. They can also fill in a text field by selecting an ``\emph{Other}'' choice. For instance, when asking about visibility of reviews, participants can respond with one of: ``\emph{Only authors and assigned reviewers can see reviews (like ICML until 2024)},'' ``\emph{Anonymized reviews released with accepted submissions; authors can opt-in for release on rejected submissions (like NeurIPS)},'' ``\emph{Anonymized reviews released with all submissions (like ICLR)},'' or ``\emph{Other}.'' We acknowledge that some combinations of policies are unreasonable (such as releasing all reviews but not all submissions).  The results are shown in \Cref{fig:conference_comparison}.

Overall, over $90\%$ of respondents prefer at least some openness in the review process. We find that $89\%$ support open review release policies. A smaller number, $33\%$, prefer releasing anonymized reviews for all submissions, and $56\%$ prefer releasing anonymized reviews for accepted submissions with author opt-in for rejected submissions. We find that $84\%$ support open commenting policies. Again, a smaller number, $38\%$, prefer allowing open, anonymized commenting immediately after manuscript submission, and $45\%$ prefer non-anonymized commenting only after conference acceptance decisions. The most-mentioned topic in the free-response question at the end of the survey is explicit, enthusiastic support for open reviewing policies (see \Cref{fig:top_free_responses}). 

These numbers are higher than the interdisciplinary responses seen by \citet{ross2017survey}, where $59\%$ support releasing review reports with articles, $55\%$ support commenting on accepted papers, and $51\%$ support community contributions to the review process (pre-decision commenting). However, the final $51\%$ figure is higher than the $38\%$ figure we receive for pre-decision commenting. In $2013$, \citet{soergel2013open} presented participants in ICLR $2013$ with the statement ``\emph{If I participate in organizing a conference in the future, I will advocate using similar open review policies}.'' They found that approximately $80\%$ of participants agreed with this statement, and $15\%$ neither agreed nor disagreed. ICLR $2013$ had the same policies as ICLR $2025$, although ICLR $2013$ only allowed non-anonymous commenting and ICLR $2025$ allowed anonymous or non-anonymous commenting. Thus, support for the fully open model was considerably higher in $2013$. Our results indicate a high level of support for partially or fully open reviewing, though the support for fully open reviewing has decreased since $2013$. \citet{soergel2013open} did not present a partially open option, so perhaps participants in that survey would have preferred a partially open model if it were presented.

Given the strong sentiment about resubmission bias, we hypothesized that those who do \emph{not} refrain from submitting to ICLR and other fully open conferences would have much higher preference rates for ICLR-style policies. Overall, this hypothesis is not borne out by the data; the results are similar for this group. However, preferences for ICLR are slightly higher among this population compared to the general population. Of this group, approximately $43\%$ prefer ICLR commenting policies, compared to approximately $38\%$ in the general population. Approximately $41\%$ prefer ICLR reviewing policies, compared to $33\%$ in the general population. Approximately $33\%$ prefer ICLR submission policies, compared to $27\%$ in the general population. However, this population still has a higher fraction of respondents who prefer partially-open policies for all three questions.

We hypothesized that area chairs and senior area chairs would be more enthusiastic about open reviewing practices. However, although they still strongly prefer open reviewing practices, there is a higher fraction of responses for closed policies among this population for all three questions, relative to the general population. Across all three categories, there is a $3$-$5$ percentage point increase in preference for closed policies.

\conferenceComparisonPlot{Responses to ``\emph{How would you set the visibility of submissions?}'', ``\emph{How would you set the visibility of anonymized reviews?}'', and ``\emph{How would you set the policy on public commenting?}''}{fig:conference_comparison}{Submissions,Reviews,Public commenting}{11.1,59.5,27.1,2.3,8.6,56.0,33.2,2.2,14.8,45.4,38.4,1.5}{0.2}{2355,2353,2352}

\paragraph{Additional Remarks}

We hope that these results will encourage further exploration.
We find that $6$ survey respondents mention the benefit of having a diversity of conference reviewing policies in the marketplace, allowing choice and experimentation. Regardless of our results, we encourage venues to continue experimenting with innovative review models. One of our respondents summed up the conversation rather nicely: ``\emph{Much of this is too complicated to be suitable for survey level analyses... this is a very complicated topic that nobody knows the right answer to yet. It will take experimentation, community work, and systemic change across stakeholders to improve the peer-review process}.''

\section{Evaluation of quality of reviews in fully open vs partially open settings}
\label{sec:llm_evaluation}

The responses we received from our survey indicate that the quality of reviews differs between open and closed reviewing settings. Specifically, 48.0\% of respondents say that openness incentivizes them to produce higher quality reviews. In this section, we take a complementary approach to evaluating the differences in reviewing quality between fully open and partially open settings. As described in further detail below, we would have ideally liked to compare reviews from open and closed venues. However, reviews from closed venues are not available publicly, which prevents us from making a direct comparison. One incidental benefit of open reviewing is that such analyses are actually possible.

We define three key metrics that serve as a measure of the quality of the review. The first is \textit{substantiation}, which is a measure of how well the review is supported with evidence and examples from the paper. Substantiation is important as it tells the authors what needs to be changed in the paper for acceptance (either in the current venue or in the future). When a reviewer substantiates the review well, it shows that they are thinking through their decision making recommendations. The second is \textit{correctness}, which measures the level of factuality in the review. Correctness is naturally a key metric, since an incorrect analysis of the paper could directly result in the wrongful acceptance or rejection of a paper. The third metric is \textit{completeness}, which is a measure of how many of the paper's contributions have been covered in the review. It is therefore an indication of whether the reviewer has missed something. To recommend the decision on a paper, it is important that the reviewer has analyzed all aspects of the paper thoroughly. Related works~\citep{vanRooyen1999RQI, superchi2019tools, goldberg2025peer} have also found that these metrics are effective ways of capturing the quality of reviews. We define these metrics in more detail later in this section.

There is evidence in prior work that AI can be used to evaluate review quality based on similar metrics~\citep{thakkar2025LLMFeedbackICLR}. We use a Large Language Model, GPT-5, as it was the latest and consistently ranked at the top of several leaderboards at the time of performing this analysis.

We evaluate reviews from ICLR 2023 (fully open) and NeurIPS 2022 (partially open). The choice of venue and year for the two settings had to satisfy the following criteria:
\begin{itemize}
    \item \textbf{Take place within similar time frames}: Reviewing styles can change year after year. Therefore, it was important to select conferences whose reviewing periods occurred close to each other.
    \item \textbf{Have similar review forms}: A more detailed review form, with a larger number of prompts and requirements, could act as a confounder when comparing the quality of reviews.
    \item \textbf{No AI contamination}: Our objective is to study how the quality of human reviews differ between the two settings. Hence, the presence of AI-generated reviews would contaminate these numbers. We could employ techniques to detect AI-generated text~\citep{tian2025gptzero, hans2024spotting} and filter out the AI-generated reviews, however, these techniques are not foolproof, and cannot guarantee control over false positives and false negatives. Thus, we only consider conferences whose review processes were completed before the launch of ChatGPT (November 2022).
    \item \textbf{Have publicly available reviews}: To analyze reviews from closed venues such as ICML, where reviews are never released, one would need to obtain permission prior to the submissions being made to the conference, which is infeasible. Prior works~\citep{kuznetsov2024what, yang2025paper} have also argued that experimentation using reviews from closed venues is difficult and researchers often consider open venues for such analyses.
\end{itemize}   

Although NeurIPS 2022 was a partially open venue, we believe that this comparison may have similar results to comparing a fully open venue with a closed venue for the following reasons. First, NeurIPS follows the closed model until decisions are released. Second, if a paper is rejected, NeurIPS lets the authors decide whether they want to make the paper together with the review publicly available. Hence, if a reviewer is of the opinion that a paper might get rejected, then they may believe that the paper will not be public, which resembles the closed model. Finally, our survey finds that a large reason reviewers modify their behaviors in open vs closed settings is due to public commenting, which is not available in NeurIPS during the review process. It is only possible to comment publicly after the decisions have been released, non-anonymously.

We only consider the accepted papers from ICLR 2023 and NeurIPS 2022 as NeurIPS releases all accepted papers and reviews, but only releases rejected papers and reviews should the authors choose to do so. Therefore, also considering rejected papers' reviews might result in biases.

We now explain the methodology behind the evaluation of the quality of these reviews and then provide the results and discuss their interpretations.

\subsection{Methodology}
\label{sec:reviewqualitymethodology}

We define the three metrics we use as a measure of the quality of the review: substantiation, correctness, and completeness. We describe the process used to evaluate a review based on each metric, using AI (specifically, Large Language Models or LLMs). We also perform robustness checks to show the validity of our methods, which can be seen in Appendix~\ref{appendix:eval_of_evals}.

\subsubsection{Substantiation}

We draw inspiration from~\citet{kennard-etal-2022-disapere} to define the following terms. An \textit{evaluative segment} is a subjective judgement of an aspect of the paper. A \textit{negative} evaluative segment is one that does not support the acceptance of the paper and a \textit{positive} evaluative segment is one that supports the acceptance of the paper. An evaluative segment in the review is \textit{well-substantiated} if the review contains appropriate evidence using examples from the paper to substantiate it. Certain segments may also be self-substantiated (these fall under well-substantiated) if they do not require any explicit support or evidence.

To evaluate the level of substantiation in the review, we first identify the evaluative segments in the review and then filter out further to select the negative evaluative segments. We only consider negative evaluative segments since a high quality review would need to elaborate well on the criticisms of the paper. Positive evaluative segments on the other hand do not necessarily need to be well elaborated. For example, statements such as ``the paper is well-written'' or ``the level of experimentation is extensive and complete'' do not require further elaboration because the evidence can be found in the paper itself. However, statements such as ``the experimentation is incomplete'' needs further elaboration in terms of what the expected output should be. To identify these segments, the LLM is provided the definitions of these terms along with a random sample of 10 data points from~\citet{kennard-etal-2022-disapere} to provide some examples of classification of segments.

The next step is to combine negative evaluative segments that cover the same aspect of the paper, but may appear in different sections of the review. To avoid cases where a large number of negative evaluative segments are identified when there are actually much fewer distinct ones, this step is performed.

Following this, we classify whether each of these segments are well-substantiated. We also classify prescriptive statements as negative evaluative. Most of these segments would be marked as substantiated, since questions/requests/suggestions by the reviewer do not normally require further elaboration, unless they are accompanied by a criticism. We pass the above instructions to the LLM along with 10 randomly chosen data points from~\citet{guo2023automatic} to provide some examples of substantiation.

For each review, the substantiation score is defined as the fraction of substantiated negative evaluative segments. The entire substantiation prompt for the LLM can be found in Appendix~\ref{appendix:prompts_used} and a sample review highlighting some negative evaluative segments along with the substantiations can be found in Section~\ref{sec:sample_review}.

\subsubsection{Correctness}
\label{subsec:correctness_reviewqualitymethodology}

We measure the level of correctness in a review as the fraction of objective segments that are factually correct. An \textit{objective segment} is defined as a segment that speaks about the content of the paper, indicates the reviewer's understanding of a part of the paper, and can be verified through the paper. We first ask the LLM to identify the objective segments in the review. We then upload the paper itself to the LLM and ask it to verify the correctness of this segment from the contents of the paper. A unique challenge encountered was that a reviewer may sometimes write in the review that ``experiment X has not been performed'', however, if we look at the paper, we do see experiment X to be a part of the paper. This does not necessarily mean that the reviewer was incorrect. In many cases, the authors add this experiment to the paper to address the reviewers' concerns. Since only the latest version of the paper is available to us, we cannot determine from the paper alone whether the objective statement is correct or not. Therefore, we also pass to the LLM all the author responses to this particular review, as well as all author responses addressed generally to all reviews. If an author has responded ``we have added experiment X to the manuscript'', we can determine that the segment was factually correct, and instruct the LLM to mark such statements as correct.

The correctness prompt can be found in Appendix~\ref{appendix:prompts_used} and a sample review highlighting some correct and incorrect objective segments can be found in Section~\ref{sec:sample_review}.

\subsubsection{Completeness}

We first define a \textit{claim} made in the paper as a contribution or output of the paper. We measure the level of completeness of the review as the fraction of all claims made in the paper that are mentioned in some form in the review. We first pass the paper to the LLM and instruct it to extract a list of all the claims made in the paper. For each claim, we ask the LLM to search the review to check whether it has been covered. We acknowledge that a review need not describe in detail every claim made in the paper, therefore, as long as the claim is covered in some form in the review, and it is clear that the reviewer has noticed this claim and its implications, we mark that claim as covered. The review forms for both ICLR 2023 and NeurIPS 2022 contain a text box which ask the reviewer to provide the summary of the review. We observe that in several reviews, the abstract of the paper is just rephrased and written in this section. Since a well-written abstract would contain all the contributions of this paper, these reviews would naturally receive a high completeness score, but this does not represent how complete the review is. Therefore, we exclude this section of the review when passing the review to the LLM.

The completeness prompt can be found in Appendix~\ref{appendix:prompts_used}.

\subsubsection{Sample review}
\label{sec:sample_review}

We present a sample review where we highlight the different categories of segments of the review as defined in our substantiation and correctness metric.

In the sample review below, the text highlighted in \classA{orange} shows the negative evaluative segments, the text highlighted in \classB{blue} show the substantiation for these segments (if present), the text highlighted in \classC{green} show the correct objective segments, and the text highlighted in \classD{yellow} show the incorrect objective segments. It can be seen that the negative evaluative segment ``\classA{The mathematical derivations are hard to follow. I think there is room for improvement in terms of clarity.}'' is unsubstantiated. The text ``\classD{Can the authors compare to more recent works such as [1]}'' is incorrect as the authors actually included this comparison in the appendix.

\begin{tcolorbox}[
  colback=white,
  colframe=blue!30,
  colbacktitle=blue!10,
  coltitle=black,
  title=\textbf{Sample Review with Highlighted Tags}
]
\small
    \classC{This paper looks at data free techniques for quantization.} \classC{The framework is to look for automorphisms on the (R,x) which the authors show is essentially power functions.} \classC{Then exponential representation are optimized to minimize reconstruction error.} \classC{Finally, experimental results are presented to show the usefulness of the work.}

    Strengths and Weaknesses:
    
    Strengths: 
    
    - \classC{The paper proposes a novel quantization technique based on power functions.} 
    
    - \classC{Some theoretical analysis is performed to describe the method.} 
    
    - Extensive experiments are performed to evaluate the method. 
    
    Weaknesses: 
    
    - \classA{The mathematical derivations are hard to follow. I think there is room for improvement in terms of clarity.} 
    
    Clarity, Quality, Novelty and Reproducibility:

    I understand this field is moving very fast and it is hard to keep track to the most recent related works. \classA{But most of the works that were compared to are already dated by a few years.} \classB{Can the authors compare to more recent works such as [1]}

    \classD{Can the authors compare to more recent works such as [1]}
    
    [1] Sakr, Charbel, et al. "Optimal Clipping and Magnitude-aware Differentiation for Improved Quantization-aware Training." International Conference on Machine Learning. PMLR, 2022.

    Summary of the review:

    The paper proposes a new quantization method, provides analyses and empirical evaluations. 

\end{tcolorbox}

\subsection{Results and discussions}

After randomly sampling one review from each accepted paper in ICLR 2023 and NeurIPS 2022, we run the review quality analysis using LLMs and compute the substantiation score, correctness score, and completeness score for each review (Section~\ref{sec:reviewqualitymethodology}). We report the mean of these scores across all reviews of each conference in Table~\ref{tab:mean_scores_and_claims} along with the mean number of claims for each metric in each review. Here, the number of claims is an umbrella term which refers to the number of negative evaluative segments in the review for substantiation, the number of objective segments in the review for correctness, and the number of claims in the paper for completeness. 

\begin{table}[h]
\centering
\begin{tabular}{|c|c|c|c|}
\hline
\textbf{Venue} & \textbf{Metric} & \textbf{Mean score} & \textbf{Mean num claims} \\
\hline
ICLR 2023 & Substantiation & 0.8711 & 4.7555 \\
NeurIPS 2022 & Substantiation & 0.8818 & 5.0556 \\
\hline
ICLR 2023 & Correctness & 0.8760 & 9.3792 \\
NeurIPS 2022 & Correctness & 0.8107 & 9.1147 \\
\hline
ICLR 2023 & Completeness & 0.4846 & 8.2876 \\
NeurIPS 2022 & Completeness & 0.4622 & 8.5257 \\
\hline
\end{tabular}
\caption{Mean scores and number of claims for each metric and for each venue. A higher mean score indicates a better score.}
\label{tab:mean_scores_and_claims}
\end{table}

We run a two-sided Fisher permutation test on each of the metrics with the test statistic being the difference in observed means between the two venues. The results can be seen in Table~\ref{tab:power_analysis_results}. Using  Holm-Bonferroni correction, with $\alpha = 0.05$, and the null for each metric being that there is no difference in the true means between the two venues, we are able to reject the null for correctness and completeness, but not for substantiation. In other words, we can conclude that there is a statistically significant difference in the mean scores of correctness and completeness between the two venues, but not for substantiation. While we acknowledge that the differences are small and that more research is required to make conclusive claims, we provide some further discussion and intuitive explanation to these numbers.

Our survey indicated that openness incentivizes higher quality reviews, but the substantiation is slightly better in NeurIPS 2022. One possible explanation is that the reviews from NeurIPS 2022 are indeed of a lower quality than the reviews from ICLR 2023. However, while writing a review, the metric that can be the most easily improved is substantiation. To achieve a higher correctness and completeness score, the reviewer would need to spend more time to understand the paper more thoroughly. The substantiation score can simply be improved using easy-to-elaborate negative evaluative segments, or even by reducing the number of negative evaluative segments. Therefore, the substantiation score is better (by a smaller margin, and with a slightly higher p-value) for NeurIPS 2022, but the correctness and completeness scores are higher for ICLR 2023.

\begin{table}[!t]
\centering
\begin{tabular}{|c|c|}
\hline
\textbf{Metric} & \textbf{$\Delta \pm$ 95\% C.I.} \\
\hline
Substantiation & $0.0107 \pm 0.0122$ \\
Correctness    & $-0.0653^{***} \pm 0.0092$ \\
Completeness   & $-0.0224^{***} \pm 0.0134$ \\
\hline
\end{tabular}
\caption{Here, $\Delta = \mu_{\text{NeurIPS}} - \mu_{\text{ICLR}}$, where $\mu_{\text{NeurIPS}}$ is the mean score of NeurIPS 2022 reviews and $\mu_{\text{ICLR}}$ is the mean score of ICLR 2023 reviews, for a given metric. Two-sided Fisher permutation tests (10,000 permutations per metric) were used to assess whether the observed mean differences are non-zero. We find that difference for substantiation has a p-value of 0.08, whereas correctness and completeness are statistically significant with p-values less than 0.001 after applying Holm-Bonferroni correction to control the family-wise error rate.}
\label{tab:power_analysis_results}
\end{table}

Intuitively, if, for example, the mean number of claims per review in a venue is 5, and we are looking at the substantiation metric, a difference of 0.02 between the mean score of NeurIPS 2022 and the mean score of ICLR 2023, indicates that in every 10 reviews, we have 1 less claim that is substantiated. In other words, one in every 10 reviewers substantiates 20\% fewer claims in that venue. 

For substantiation, to calculate the difference in the number of unsubstantiated claims between the two venues, on average, for each review, we perform the following using the values in Table~\ref{tab:mean_scores_and_claims}. We multiply the mean fraction of unsubstantiated claims with the mean number of claims for each venue, and find the difference between them. We find that approximately for 1 in every 65 reviews, ICLR 2023 has one more unsubstantiated claim, or, approximately 92 reviews in ICLR 2023 have one more unsubstantiated claim than NeurIPS 2022.

Similarly, for correctness, 1 in every 1.78 reviews of NeurIPS 2022 has one more incorrect claim than ICLR 2023, or, approximately 5494 reviews in NeurIPS 2022 have one more incorrect claim than ICLR 2023. As mentioned in Section~\ref{subsec:correctness_reviewqualitymethodology}, we pass all author responses to the review together with the paper and the review itself to the LLM when evaluating the correctness metric. This is because only the latest version of the paper is available on OpenReview. If the author responds to the reviewer mentioning they have addressed any points made by the reviewer and modified the paper, the LLM does not mark these points as incorrect. For ICLR 2023, the mean number of author responses to each review is 2.26 (for 1573 papers) and for NeurIPS 2022, is 2.04 (for 2671 papers). Further, the discussion period for ICLR 2023 was about six weeks. The authors could still update their papers for the first two weeks and continue responding to reviewers/area chairs for the next four weeks. For NeurIPS 2022, the discussion period was only two weeks in total. Therefore, the increased discussion in ICLR 2023 could have led to the authors stating the changes they have made in the paper more clearly due to which the LLM may have marked a higher fraction of claims as correct. This is an important point to note while considering the difference in means of correctness scores between the two venues.

To summarize, the differences in scores for each metric and their interpretations are preliminary results from using LLMs to evaluate the quality of reviews, and further analysis should be done to confirm these differences.

We also analyze the number of reviews in each venue that have at least one unsubstantiated claim, and the number of reviews in each venue that have at least one incorrect claim. The results can be found in Table~\ref{tab:atleast_one_unsub}. They points to the fact that reviews that are perfectly substantiated have a lower number of claims, thereby making them easier to substantiate. 

\begin{table}[!t]
\centering
\begin{tabular}{|c|c|c|c|c|}
\hline
\textbf{Venue} & \textbf{Metric} &
\makecell{\textbf{\% reviews with}\\\textbf{score < 1.0}} &
\makecell{\textbf{Mean no. claims}\\\textbf{given score < 1.0}} &
\makecell{\textbf{Mean no. claims}\\\textbf{given score = 1.0}} \\
\hline
ICLR 2023 & Substantiation & 40.96\% & 5.58 & 4.18 \\
NeurIPS 2022 & Substantiation & 40.37\% & 5.76 & 4.57 \\
\hline
ICLR 2023 & Correctness & 60.57\% & 9.86 & 8.64 \\
NeurIPS 2022 & Correctness & 77.87\% & 9.51 & 7.72 \\
\hline
\end{tabular}
\caption{Computing the percentage of reviews with at least one unsubstantiated claim, and the percentage of reviews with at least one incorrect claim, along with the mean number of claims for each metric conditioned on the score being a perfect 1.0, for each venue.}
\label{tab:atleast_one_unsub}
\end{table}

We also find in Table~\ref{tab:correlation_newmetrics} that there is very low correlation between each of these three metrics and the length of the review, the confidence of the reviewer (obtained through the reviewer confidence score in the review), and the score given by the reviewer (obtained through the review score in the review). These correlations are not statistically significant as they have high p-values. We note that another experiment conducted in a top-tier ML/AI conference, involving manual annotations of reviews, also found no correlation between the quality of reviews and the reviewer's self-reported confidence~\cite[Section 10.1.2]{shah2022challenges}.

\begin{table}[!t]
\centering
\begin{tabular}{|c|c|c|c|}
\hline
\textbf{Metric 1} & \textbf{Metric 2} & \textbf{Pearson's \(r\)} & \textbf{Spearman's \(\rho\)} \\
\hline
Substantiation & Review length & 0.1611 & 0.0927  \\
Substantiation & Reviewer confidence & 0.0539 & 0.0187 \\
Substantiation & Reviewer score & -0.0109 & 0.0455 \\
\hline
Correctness & Review length & 0.3023 & 0.3256  \\
Correctness & Reviewer confidence & 0.1617 & 0.1500 \\
Correctness & Reviewer score & 0.0690 & 0.0562 \\
\hline
Completeness & Review length & -0.1151 & -0.1191  \\
Completeness & Reviewer confidence & -0.0511 & -0.0479 \\
Completeness & Reviewer score & 0.1822 & 0.1719 \\
\hline
\end{tabular}
\caption{Pairwise Pearson's \(r\) and Spearman's \(\rho\) coefficients between each of the three metrics (substantiation, correctness, completeness), against review length, reviewer confidence, and reviewer score across both venues (n=4334).}
\label{tab:correlation_newmetrics}
\end{table}

\section{Related work}
\label{sec:related_work}

We discuss other surveys of attitudes to open reviewing practices in \Cref{subsec:other_surveys}, and other analyses of reviewer behavior and review quality in open venues in \Cref{subsec:quality_in_open}.

\subsection{Surveys of attitudes to open peer review}
\label{subsec:other_surveys}

\citet{soergel2013open} explored the landscape of peer-review policies, described the then novel OpenReview.net platform, and piloted the system with the fully open $2013$ International Conference on Learning Representations. The present survey follows up on many of the same questions, and we have included comparisons with the older survey results throughout our analysis.

\citet{charness2022improving} studied the state of peer review in economics, surveying over $1{,}400$ economics researchers over the course of a year. We survey approximately $1.7$ times as many researchers, though our population is centered around those who participated in the ICLR $2025$ conference. They listed a set of proposals and asked: ``\emph{Below is a list of proposals to improve peer reviews. On a scale from 1 to 5, how useful do you find each of them?}'' They found that among their respondents, approximately $40\%$ were favorable or very favorable towards ``\emph{Making the history of (anonymous) reviews and authors’ responses publicly available},'' with a further $20\%$ expressing a neutral position. When asked about ``\emph{Making all reports available to all of the reviewers and making sure reviewers know this is being done},'' nearly $70\%$ were favorable or very favorable, and a further $10-15\%$ were neutral. A separate question asked ``\emph{Another recent trend is to make the history of reports/responses to referees publicly available in an anonymized way unless the reviewers choose to disclose their identity; see e.g., Nature Communcations. On a scale from 1 to 5, how favorable would you be to such a policy?}'' Approximately $50\%$ stated that they were favorable or very favorable towards the idea, with a further $25\%$ expressing a neutral position. The survey authors caution that differences in wording may be the source of differences in results, but overall the results show that economists are largely favorable of releasing anonymized review reports along with author responses.

\citet{ross2017survey} collected over $3{,}000$ survey responses across many disciplines, though they note that $90\%$ of responses come from researchers in science, technology, and medicine. Overall, they find that approximately $60\%$ of respondents are in favor of some openness in the peer-review process, though open identity is strongly disfavored. Participants strongly supported ``open interaction, open reports, and final-version commenting'' in particular. They found that computer science and IT researchers were among the more supportive disciplines for open reviewing practices. Some of their results largely agree with ours; a large majority of respondents in their survey said that open review reports provide useful information to the reader, and $46\%$ said that publishing review reports would make reviewers less likely to make strong criticisms (vs. $32\%$ who disagreed). They also found that $60\%$ of respondents agreed that publishing review reports would increase the overall quality of peer review. They found that only $27\%$ agreed with the statement that potential authors are less likely to submit to  venues that publish reviews, while we found that about the same number of authors report \emph{having refrained in this way}. This qualifies this result in a rather remarkable way, showing that although a large majority of participants do not refrain from submitting and do not think others do, a significant fraction do indeed refrain.

\citet{besanccon2020open} ran a survey on $30$ researchers in the field of human-computer interaction (HCI), who had participated in the alt.chi peer-review track of the CHI conference. This track utilizes a fully open reviewing process, in which reviewers' non-anonymized review reports are released with all papers. However, at the end of this process, the system is closed. Some authors are requested to attach the reviews and discussions for their papers to their final manuscripts, but others are not re-released.  Respondents were overall very favorable to the process, and highlighted increased transparency, constructiveness of reviewers' reports, and the benefits of having public discussions about papers. Participants highlighted drawbacks that are echoed in our feedback, such as the potential for public commenting during the reviewing process to make reviewing more akin to a ``popularity contest.''

\subsection{Reviewer behavior in open reviewing venues and review quality}
\label{subsec:quality_in_open}

\citet{bravo2019effect} studied the effects of reviewers' behavior in five different journals when they were told that their reviews would be made anonymously publicly available. They observed that only 35.8\% of invited reviewers were willing to review, more submissions were rejected or were given major revisions, reviewers who wrote positive reviews were more keen to reveal their identity, and the polarity of the reviews remained the same. \citet{stelmakh2021prior} show that reviewers become negatively biased when they learn that a paper is a resubmission, a phenomenon common in open reviewing venues.

\citet{van2019think} argues that open peer review serves as a crucial educational tool for PhD students. By making reports public, students can learn what constitutes a high quality review and understand the decision-making standard. They also suggest that open identities allow PhD students, who often ``ghostwrite'' reviews for PIs, to finally receive visible credit for their labor, and that the public nature of the process may encourage a more constructive tone from senior reviewers.

\citet{tran2020open} performed a quantitative analysis of ICLR data (2017 to 2020) to evaluate the efficacy of the open process. They examined the discussion phase and found that while mean review scores tend to increase during the rebuttal period, the process still suffers from significant ``institutional bias'' and a lack of reproducibility. They argue that while openness provides transparency, it does not automatically solve issues of fairness or bias in the decision-making process.

\citet{vanRooyen1999RQI} was among the first works to propose and evaluate criteria or metrics to evaluate the quality of peer reviews. They suggest metrics such as whether the reviewer discussed the importance of the research, whether the reviewer substantiated their comments, whether the reviewer's comments were constructive, and whether the reviewer was comprehensive. \citet{superchi2019tools} systematically review all tools that have been used to assess peer-review quality. Their analysis suggests that clarity, constructiveness, thoroughness, fairness, knowledgeability, and tone are common and effective ways to capture it. \citet{goldberg2025peer} use understanding, coverage, substantiation, and constructiveness to evaluate the quality of reviews.

Several works introduce datasets annotating reviews based on various aspects or annotating sentence-level text which can be applied to peer review. \citet{ghosal2022peer} introduce a dataset of approximately 1,200 reviews, each manually annotated at the sentence level and given four tags based on four criteria; which section of the paper it corresponds to, what aspect (e.g., novelty, soundness, originality), tone (positive, negative, summary), and importance. \citet{kennard-etal-2022-disapere} create a dataset DISAPERE, which consists of 20k sentences, each with multiple categories, such as action(evaluative, request), aspect (substance, clarity), polarity, etc. \citet{guo2023automatic} publish a dataset to automate the evaluation of substantiation in peer reviews. The dataset, SubstanReview, consisting of full-length reviews, each containing annotations of evaluative segments in the review along with its substantiation if present. \citet{bharti2022betterpr} introduce a dataset of English sentences containing annotations of whether it is constructive or not.

\section{Limitations and Conclusion}
\label{sec:limitations_and_conc}

We first discuss limitations of the survey. The survey was sent via an email by the program chair of the ICLR 2025 conference to participants of that conference. Since this is a fully open reviewing venue, this population may have a preference for open reviewing policies. It is important to note, however, most machine learning researchers regularly submit to multiple venues that also include comparable venues with partially open and closed reviewing policies. Thus the set of researchers who participate in ICLR would naturally overlap substantially with other major machine learning venues. 

We now estimate our response rate. There were $11{,}603$ submissions to ICLR $2025$, $18{,}325$ reviewers, $823$ area chairs, and $71$ senior area chairs \citep{ICLR2025FactSheet}. These participation numbers suggest a response rate of $11.2\%$ for reviewers, $38.6\%$ for area chairs, and $88.7\%$ for senior area chairs. If we assume that there are approximately $3$ unique authors per paper, as was the case in the ICML $2023$ conference \citep{su2024analysis}, then the author response rate is approximately $6\%$. 

We did not offer compensation to respondents. There may have been self-selection bias since our survey was distributed via an anonymous link over email; our respondents are likely to be the subset of ICLR $2025$ participants who are most engaged or who have stronger opinions about open reviewing practices. Any survey can only represent self-reported experiences and perceptions, rather than estimating outcomes directly via controlled experiments or observation. Our survey also only captures a single moment in time. 

We now discuss limitations of our annotation study. We are not able to obtain closed review data for our annotation study, which would be the most reliable data to test the effects of openness on review quality. In particular, we are unable to obtain reviews of most rejected papers for NeurIPS $2022$. Although we perform robustness checks for the AI annotation procedure, these robustness checks were done over a fairly small annotation set for completeness and correctness (substantiation was tested against an existing benchmark). We also only annotate two reviewing venues. Expanding the annotation process to more venues would bolster the validity of our findings.

To conclude, we survey $2{,}385$ respondents who have recently participated in top-tier peer-review venues in machine learning and computer science. We find that this community is highly positive about open peer-review practices, with over $80\%$ of respondents supporting release of reviews for accepted papers and allowing public commenting on accepted papers. However, fewer participants ($27.1\%$) support releasing rejected manuscripts. We find that the biggest perceived benefits are improvements to public discussions and understanding, making the decision process  more fair, and incentivizing reviewers to exert effort. A large number of respondents are concerned with resubmission bias. This fact aligns with the relatively low fraction of respondents who preferred releasing rejected manuscripts during the review process. Participants also believe that open reviewing practices can help reduce AI slop reviews and manuscript submissions. Our automated review annotation procedure shows marginally improved review quality in a fully open venue (ICLR $2023$) compared to a partially open venue (NeurIPS $2022$), though more work is needed to fully contextualize this result. 

We hope these results help conferences and journals make more informed decisions about their reviewing policies.

\section*{Acknowledgments}
Our deepest thanks go to those who thoughtfully completed the survey. The responses we received were incredibly informative, helping inform policies for peer reviewing in our community. We also profusely thank Carl Vondrick, the Senior Program Chair of ICLR $2025$, for sharing the link to the survey with the  participants in ICLR $2025$. We are grateful to our friends and collaborators who piloted the survey and gave us useful feedback: Ziming Luo, Yair Zick, Vignesh Viswanathan, Dmitry Petrov, Alexander Goldberg, and Nicholas Perello. Thank you to Ameet Talwalkar for early conversations about the public review analysis. 

This work was supported by NSF 2200410 and 1942124 and ONR N000142512346.

\bibliographystyle{plainnat}
\bibliography{abb, main}

\newpage
\appendix

~\\~\\\noindent{\bf \Large Appendices}

\section{Survey Pilot}
\label{appx:survey_pilot}

We sent an early version of the survey to $15$-$20$ researchers in machine learning and computer science, and received comments from $6$ of them. We asked the following questions:

\begin{itemize}
\item Do you find the initial description of the review policies clear?
\item Were any of the questions ambiguous or confusing?
\item Did you feel that an answer choice was missing from any question, such that you could not express your true opinion in response to the question? 
\item Is there some important question you feel is left off the survey?
\item Was the section on selecting the set of policies you would recommend to future conference organizers confusing?
\end{itemize}

One other question we aimed to address with our pilot was which items should be included in the ranking question (``\emph{Consider the potential impacts of fully open reviewing policies that release rejected submissions and reviews. Please rank these outcomes based on how much they matter to you overall—taking into account both how likely each one seems to you, and how good or bad it would be if it happened. Rank from 1 (matters most) to 5 (matters least). Use each rank exactly once.}''). At this time, we included $7$ items to rank. These included the $5$ items that are in the final survey, as well as the items ``\emph{Important perspectives from reviewers and commenters are shared with the general public}'' and ``\emph{Novice scientists learn how to write reviews}.'' We removed these two items since they had consistently low ranks from our pilot participants, and we received feedback that there were too many items to rank.

\section{Prompts Used to Evaluate Review Quality}
\label{appendix:prompts_used}

\begin{tcolorbox}[
  breakable,
  toprule at break=0pt,
  bottomrule at break=0pt,
  colback=white,
  colframe=blue!30,
  colbacktitle=blue!10,
  coltitle=black,
  title=\textbf{Substantiation Prompt},
]
\small
    SEGMENT CLASSIFICATION DATA:
    
    <10 samples points from~\citet{kennard-etal-2022-disapere}>
    
    SUBSTANTIATION DATA:
    
    <10 sample points from~\citet{guo2023automatic}>
    
    Here is a review of a scientific paper:
    
    <Review of paper>
    
    Your goal is to identify the substantiated and unsubstantiated segments from the review and provide them. Follow the below instructions to complete the task; For steps 1 and 2, make use of the SEGMENT CLASSIFICATION DATA in the beginning to understand how to classify segments.
    
    1. Identify the evaluative segments in the review. An evaluative segment is defined as a subjective judgement of an aspect of the paper. They can be sentences, multiple sentences, or even part of a sentence. Sometimes this may also be in the form of a question.
    
    2. For each evaluative segment identified in step 1, identify the negative evaluative segments. A negative evaluative segment is an evaluative segment that does not support the acceptance of the paper.
    
    3. After identifying all negative evaluative segments, review them carefully. If you see segments that continue the same line of criticism where the second segment provides examples, explanations, or elaborations of the first (especially when connected by phrases like "In particular," "For instance," etc.), then this second segment is usually a substantiation. Therefore, if you feel it is more a substantiation, use this segment as a substantiation as per step 4. Otherwise, merge these into a single segment before proceeding to the substantiation analysis. Note that often times we tend to classify the substantiation also as a negative evaluative segment. Do not make that mistake.
    
    4. For this step, make use of the SUBSTANTIATION DATA in the beginning. For each negative evaluative segment identified in step 2, identify whether it has been well-substantiated or not. A negative evaluative segment is well-substantiated if the review contains appropriate evidence using examples from the paper to substantiate it, i.e., the substantiation should elaborate on the negative evaluative segment. If you identify the negative evaluative segment as well-substantiated, provide the segment along with its substantiation. Some points to note:
    \begin{itemize}
        \item It may happen that the part of the review that substantiates the negative evaluative segment is far away from the segment itself; so check the entire review when you are looking for a segment's substantiation. 
        \item It may happen that the segment is self-substantiated, i.e., the substantiation is within the segment itself; in such cases, say that it is self-substantiated. The substantiation in such cases MUST be a substring of the segment itself, be sure to double check that.
        \item There may be segments that are prescriptive, i.e., they provide the authors with changes or suggestions to be made in the paper. These could be editorial suggestions such as terminologies, grammar, writing style, etc., or they may be any conceptual suggestions. These segments are generally self-substantiated because they don't require evidence/support as reviewers should make suggestions to improve the paper.
        \item In cases where the segments are questions, thoughts, requests, clarifications, etc. where the reviewer is talking about a specific aspect of the paper, these segments would normally be self-substantiated. Only in extreme cases, where the reviewer is extremely vague, and no other part of the review elaborates on this segment, it would be unsubstantiated.
        \item If you identify the negative evaluative segment as not substantiated, provide the segment along with the text `Not substantiated'.
        \item The substantiation you provide must be directly taken from the review, do not provide any explanation, only reproduce the part of the review.
    \end{itemize}

    Please provide you response step-wise as described.
    
    After the step-wise response, output a list of the format: [(`<negative evaluative segment 1>', `<substantiation, if any>'), (`<negative evaluative segment 2>', `<substantiation, if any>'), ...], for each of the negative evaluative segments identified. If there is no substantiation, leave the substantiation as an empty string (like this: `'). If it is self-substantiated, just add the negative evaluative segment again as the substantiation.
\end{tcolorbox}

\begin{tcolorbox}[
  colback=white,
  colframe=blue!30,
  colbacktitle=blue!10,
  coltitle=black,
  title=\textbf{Correctness Prompt}
]
\small
    I am trying to study the quality of peer reviews using correctness as a metric. Your job is to follow the below instructions to determine the level of correctness in the given review. You are provided the manuscript along with its review. You are also provided the author replies to this particular review as well as the common replies by the authors.
    
    1. Identify the segments in the review that are OBJECTIVE in nature and speak about the contents of the paper. They can be sentences, multiple sentences, or even part of a sentence. These segments would indicate the reviewer's understanding (or interpretation) of a part of the paper and may be present as a part of the summary, strengths, weaknesses, suggestions, or anywhere else. 
    
    2. For each segment, look through the paper, and determine whether it is correct or not. For a segment to be incorrect, the reviewer's interpretation should be different from what the author is conveying. If the author is ambiguous, then the reviewer's interpretation can be considered as correct. If a segment is correct, the facts should be verifiable from the paper itself. If a segment is incorrect, there should be excerpts from the paper which can be used to show that the segment is incorrect.
    
    3. The manuscript provided to you may be an updated version addressing the review's critiques. For example, if the review says "Experiment X has not been performed" but you see that the manuscript does contain "Experiment X", it may be because the author incorporated the reviewer's comments and updated the manuscript. That is why you need to determine what the state of the original manuscript was using the replies to the reviews. In the example here, if you see in a rebuttal that "Experiment X" was added to the manuscript, or if you see another review that also mentions "Experiment X" is absent, then you can conclude that in the original version of the paper, i.e., the version of the paper the review was written on, "Experiment X" was actually absent, and hence the segment is actually correct.
    
    4. Therefore, check all the replies to the review for each objective statement, and if the reply contains something like "Experiment X has been added" or "Baselines have been updated to include Y", then the objective statement which states "Missing experiment X" or "baselines do not include Y" are in fact correct because they were comments made on the original paper.
    
    5. If it is an incorrect segment, you will have to provide the reasoning behind why it is incorrect alongside the segment. If it is a correct segment, just provide an empty string alongside the segment, as per the provided format.
    
    Provide your answer in the following format:
    
    After analysis, add the text ``final\_answer:'' AND THEN IT SHOULD BE followed by: [(``<evaluative segment>'', ``<reasoning if incorrect>''), (``<evaluative segment>'', ``''), ...]
    
    Here is the review of a scientific paper:
    <Review>
    
    Here are all the replies to this review:
    <Replies>
\end{tcolorbox}

\begin{tcolorbox}[
  colback=white,
  colframe=blue!30,
  colbacktitle=blue!10,
  coltitle=black,
  title=\textbf{Completeness Prompt}
]
\small
    I am trying to study the quality of peer reviews using completeness as a metric. Your job is to follow the below instructions to determine the level of completeness in the given review. You are provided the manuscript along with its review.
    
    1. Make a list of all the claims made in the paper. These claims must all be part of the contributions of the paper.
    
    2. For each claim, search the review of the paper to see if it has been covered or not. It need not have been covered in detail, but as long as it is clear that the reviewer has noticed this claim and its implications, it counts as being covered. 
    
    3. If a claim has been covered in the review, mark it as 1 and if it has not been covered in the review, mark it as 0.
    
    Provide your answer in the following format:
    
    After analysis, add the text ``final\_answer:'' AND THEN IT SHOULD BE followed by: [(<claim>, 0 or 1), (<claim>, 0 or 1), ...]
    
    Here is the review of a scientific paper:
    <Review>
\end{tcolorbox}

\section{Power Analysis} 
\label{sec:power_analysis}

To determine if we can reliably detect a difference in the quality, through each metric, between the two venues, we ran a two-sided Fisher permutation test on the absolute mean difference
\[
T_{\text{obs}} \;=\; \bigl|\, \bar{B} - \bar{A} \,\bigr|.
\]
where $A$ and $B$ are scores from the two venues (ICLR 2023 and NeurIPS 2022), $\bar{A}$ and $\bar{B}$ are their sample means, and $T_{\text{obs}}$ is the observed absolute mean difference.

Under the null hypothesis that ICLR and NeurIPS reviews for that metric come from the same distribution, we pool all scores, repeatedly shuffle the labels while keeping the group sizes fixed at $(n_{\text{ICLR}}, n_{\text{NeurIPS}})$, and recompute the same statistic on each shuffled dataset. The permutation p-value is then the fraction of shuffled statistics that are at least as large as $T_{\text{obs}}$, with a small $+1$ correction:
\[
p \;=\; \frac{\#\{ T_{\text{perm}} \ge T_{\text{obs}} \} + 1}{\text{(\# permutations)} + 1}.
\]
Here $T_{\text{perm}}$ is the same statistic computed after a random relabeling, \# permutations is the total number of random relabelings (we use $10{,}000$); the $+1$ avoids zero p-values with a finite number of permutations.

Because we test three metrics on the same two venues, we correct for multiple testing to keep the family-wise error rate at $0.05$. Each metric has the same null/alternative,
\[
H_{0,m}: \mu^{(m)}_{\text{NeurIPS}} = \mu^{(m)}_{\text{ICLR}}
\quad \text{vs.} \quad
H_{1,m}: \mu^{(m)}_{\text{NeurIPS}} \neq \mu^{(m)}_{\text{ICLR}}, \qquad m=1,2,3,
\]
where $m\in\{1,2,3\}$ indexes the three metrics and $\mu^{(m)}_{\text{ICLR}}$ and $\mu^{(m)}_{\text{NeurIPS}}$ denote the population means for metric $m$.

At analysis time we report Holm-Bonferroni results. Here, we sort the three permutation p-values and test them at the step-down thresholds $\{0.05/3,\,0.05/2,\,0.05\}$ (i.e., $\alpha/3,\ \alpha/2,\ \alpha$ with $\alpha=0.05$).

\section{Evaluation of Evaluations}
\label{appendix:eval_of_evals}

To verify the ability of LLMs to correctly evaluate the quality of reviews, we consider a mixture of approaches. For substantiation, we evaluate the LLM responses on the SubstanReview dataset~\citep{guo2023automatic}. Since, to the best of our knowledge, there are no datasets available that suit our requirements, to benchmark the completeness and correctness metric, we rely on human annotations. We also performed annotations for the substantiation metric.

\subsection{Benchmarking on an Existing Dataset}
\label{appendix:substanreview_benchmark}

To evaluate the LLM responses for the substantiation metric, we use the SubstanReview dataset~\citep{guo2023automatic}. The test set of SubstanReview consists of 110 reviews, each of which contain word-level classifications of evaluative segments along with their substantiation, if any. We consider only the negative evaluative segments and use multiple metrics, the word-level precision, recall, and F1 scores, Krippendorff's alpha score, the Spearman r correlation and the Pearson rho correlation. 

\subsubsection{Word-Level F1 Scores}
\label{substanreviewf1}

For the word-level precision, recall, and F1 scores, we map both the ground-truth and model phrases back to the original review text. Concretely, we first clean the text to handle real-world messiness (such as line-break hyphenation like ``com- plementary'', inconsistent whitespace, etc.). We then locate claims and substantiations with a context-aware matcher that’s tolerant to punctuation/spacing and ellipses. This makes sure we only give credit when the model finds the right words in the right place (same span in the review), not just anywhere the words might also appear.

Once phrases are aligned to spans, we label words inside those spans and compute micro-averaged precision, recall, and F1 at the word level for (i) negative evaluative claims and (ii) their substantiation. We observe the results in Table~\ref{tab:substanreview_f1scores}. We note that while the F1 score for substantiation outperforms the F1 score for the best model of \citet{guo2023automatic}, it still does not achieve high precision. On observing individual cases closely for why this is the case, we see that the LLM picks out many more negative evaluative segments than the dataset does, which are indeed negative evaluative segments, and also picks out questions/requests/suggestions that have a negative tone, whereas the dataset does not. This is because our understanding of evaluative segments is much broader than the definitions provided for the SubstanReview dataset, which states that evaluative segments are ``subjective statements that reflect the reviewer’s evaluation of the paper. Subjectivity can be expressed either explicitly by descriptions of the writer’s mental state or implicitly through opinionated comments.''. Due to this, the precision for the claims is low, and by propagation of error, the precision for the substantiation is low as well. We note however that the recall is extremely high as the model always picks out the negative evaluative claim and its substantiation as noted by the dataset.

\begin{table}[h]
\centering
\begin{tabular}{|c|ccc|ccc|}
\hline
 & \multicolumn{3}{c|}{\textbf{Claim}} & \multicolumn{3}{c|}{\textbf{Substantiation}} \\
\cline{2-7}
 & Precision & Recall & F1 & Precision & Recall & F1 \\
\hline
LLM used (GPT-5) & 0.1741 & \textbf{1.0000} & 0.2965 & 0.3821 & \textbf{0.8347} & \textbf{0.5242} \\
Best model from~\citet{guo2023automatic} & \textbf{0.8143} & 0.6748 & \textbf{0.7356} & \textbf{0.7502} & 0.3306 & 0.4523 \\
\hline
\end{tabular}
\caption{Comparison of the word-level precision, recall, and f1 scores for identifying the claim and substantiation tags on the SubstanReview dataset~\citep{guo2023automatic} by the LLM we used (GPT-5) and the best model from their paper. }
\label{tab:substanreview_f1scores}
\end{table}

\subsubsection{Krippendorff's \texorpdfstring{\(\alpha\)}{alpha} Score}

Krippendorff's alpha answers the question ``how often does the model put the same tags in the same places as the dataset, above chance?'' We turn each review into minimal units (words) and tag every word as one of three categories: claim, substantiation, or neither, for both the ground truth annotations and the LLM. Krippendorff's alpha then summarizes agreement as
\[
\alpha = 1 - \frac{D_o}{D_e},
\]
where \(D_o\) is the observed word-level disagreement (i.e., the fraction of words whose labels do not match between the LLM output and the dataset), and \(D_e\) is the disagreement expected by chance from the pooled label frequencies. High \(\alpha\) means the model is matching the dataset's segmentation and the LLM's segmentation in the same spans, not just getting overall counts right. To handle messy real-world text again, we map phrases back to the original review with a context-aware matcher that tolerates punctuation, ellipses, line-break hyphenation, inconsistent whitespace, etc.

We find the global Krippendorff's \(\alpha\) (over all words over all reviews) to be 0.33 and the macro \(\alpha\) averaged across reviews to be 0.34. While these scores again appear lower, on examining the reviews case by case, we attribute to similar reasons explain in Section~\ref{substanreviewf1}.

\subsubsection{Pearson and Spearman Correlation}

For a review-level summary, we utilize the substantiation score for each review, i.e., the fraction of negative evaluative claims that are substantiated. Following standard practice to avoid division-by-zero and to reflect that the absence of negative claims implies nothing to substantiate~\citep{guo2023automatic}, we set the rate to \(1.0\) when a review contains zero negative evaluative claims. We then compute both Pearson's \(r\) and Spearman's \(\rho\) between the LLM's rates and the dataset's rates across reviews. We observe that Pearson's \(r\) is 0.1724 with a p-value of 0.0744 and Spearman's \(\rho\) is 0.1377 with a p-value of 0.155. The correlations are low, indicating that even when word-level recall is high (Section~\ref{substanreviewf1}), the model and the dataset often disagree on which reviews are well-substantiated overall.

To understand this low correlation, we analyze reviews by looking at the absolute substantiation–rate gap. The largest gaps reveal that the LLM often marks certain claims as substantiated when the ground truth dataset does not (we believe in these cases that they are indeed well-substantiated), and the LLM extracts many more negative evaluative claims as explain in Section~\ref{substanreviewf1}. There are often cases when the dataset identifies no negative evaluative claims at all whereas the LLM does.

\subsection{Annotations}
\label{appendix:annotations}

To perform annotations on all the metrics; completeness, correctness, and substantiation. Specifically, we randomly sampled five reviews each from ICLR 2023 and NeurIPS 2022 and asked two annotators to follow the exact instructions provided to the LLM (Section~\ref{sec:reviewqualitymethodology}) and note the correctness score, the completeness score, and the substantiation score for each of these reviews. For the correctness metric, the LLM was first asked to pick out objective segments in the review, for the completeness metric, the LLM was first asked to extract all the claims made in the paper, and for the substantiation metric, the LLM was first asked to identify all the negative evaluative claims in the review. We provide these objective segments and claims as starting points for the annotators. The annotators then marked each of them as correct/incorrect, covered/not covered, substantiated/unsubstantiated respectively. We then compute Pearson's \(r\) and Spearman's \(\rho\) between each of the LLM, annotator\_1 and annotator\_2.

We compute Pearson's \(r\) and Spearman's \(\rho\) between each of the LLM, annotator\_1 and annotator\_2. For the completeness metric, we find the scores to be fairly correlated. For the correctness and substantiation metrics, we find the correlations to still be positive but not as high. On discussion for each review, we find that in the cases where there is a large difference in scores, the LLM tends to be more accurate than the annotators. We also note that since there are only 10 samples, one outlier contributes largely to the correlation coefficients. The complete correlation results can be found in Table~\ref{tab:correlation_annotations}.

We also calculate the mean difference in scores between ICLR 2023 reviews and NeurIPS 2022 reviews and find that for the correctness and completeness metric, this difference is similar for the LLM, annotator\_1 and annotator\_2. This can be seen in Table~\ref{tab:delta_means_annotations}. It should be noted that since these are only across five reviews from each venue, the values themselves are not representative of the actual values observed when we calculate the means across all reviews. These numbers are just sanity checks that the LLM is working as expected. For the substantiation metric, the numbers are quite different, which is why benchmarking on the SubstanReview dataset (Section~\ref{appendix:substanreview_benchmark}) gives us a better sanity check.

\begin{table}[h]
\centering
\begin{tabular}{|c|c|c|c|c|}
\hline
\textbf{Metric} & \textbf{Evaluator 1} & \textbf{Evaluator 2} & \textbf{Pearson's \(r\)} & \textbf{Spearman's \(\rho\)} \\
\hline
Correctness & Annotator\_1 & LLM & 0.3776 & 0.3165 \\
Correctness & Annotator\_1 & Annotator\_2 & 0.4301 & 0.6492 \\
Correctness & LLM & Annotator\_2 & 0.4114 & 0.1030 \\
\hline
Completeness & Annotator\_1 & LLM & 0.8709 & 0.8632  \\
Completeness & Annotator\_1 & Annotator\_2 & 0.7334 & 0.6585 \\
Completeness & LLM & Annotator\_2 & 0.5913 & 0.6201 \\
\hline
Substantiation & Annotator\_1 & LLM & 0.1286 & 0.3106  \\
Substantiation & Annotator\_1 & Annotator\_2 & 0.1730 & 0.5797 \\
Substantiation & LLM & Annotator\_2 & 0.6553 & 0.6157 \\
\hline
\end{tabular}
\caption{Pair-wise Pearson's \(r\) and Spearman's \(\rho\) coefficients between the two annotators and the LLM for the correctness, completeness, and substantiation metric, across five reviews each from ICLR 2023 and NeurIPS 2022 (n=10).}
\label{tab:correlation_annotations}
\end{table}

\begin{table}[h]
\centering
\begin{tabular}{|c|c|c|}
\hline
\textbf{Evaluator} & \textbf{Metric} & \textbf{$\Delta$} \\
\hline
Correctness & Annotator\_1 & 0.0352 \\
Correctness & LLM & 0.1053 \\
Correctness & Annotator\_2 & 0.0818 \\
\hline
Completeness & Annotator\_1 & 0.2480 \\
Completeness & LLM & 0.2051 \\
Completeness & Annotator\_2 & 0.2121 \\
\hline
Substantiation & Annotator\_1 & -0.0286 \\
Substantiation & LLM & 0.3893 \\
Substantiation & Annotator\_2 & 0.1429 \\
\hline
\end{tabular}
\caption{Difference in observed mean scores across venues reported by each evaluator, where $\Delta$ is the difference between the mean score for NeurIPS 2022 and the mean score for ICLR 2023 (n=10).}
\label{tab:delta_means_annotations}
\end{table}

\end{document}